\documentclass[sigconf, nonacm]{acmart}

\AtBeginDocument{%
  \providecommand\BibTeX{{%
    \normalfont B\kern-0.5em{\scshape i\kern-0.25em b}\kern-0.8em\TeX}}}

\usepackage{graphicx}
\usepackage{svg}
\usepackage{caption}
\usepackage{subcaption}
\usepackage{booktabs}
\usepackage{geometry}
\usepackage{array}
\usepackage{multirow} 
\usepackage{tabularx}
\usepackage{enumitem}
\usepackage{xcolor}
\usepackage{soul}

\newcommand{\rv}[1]{\textcolor{black}{#1}}

\begin{document}

\title[Metaphors for Governing Language Model Data for Creative Writing]{Seed Bank, Co-op, Stoop Swap: Metaphors for Governing Language Model Data for Creative Writing
}


\author{Alicia Guo}
\email{axguo@cs.washington.edu}
\affiliation{%
  \institution{University of Washington}
  \city{Seattle}
  \country{United States}
}

\author{Carly Schnitzler}
\email{cschnit1@jh.edu}
\affiliation{%
  \institution{Johns Hopkins University}
  \city{Baltimore}
  \country{United States}
}

\author{Katy Gero}
\email{katy.gero@sydney.edu.au}
\affiliation{%
  \institution{University of Sydney}
  \city{Sydney}
  \country{Australia}
}

\renewcommand{\shortauthors}{Guo et al.}



\begin{abstract}
How might we govern a language model run for and by creative writers? While generative AI use is on the rise, many language models are created and owned in ways that limit writers’ consent, participation, and control. We report on four workshops where over one hundred creative writers came up with and analyzed metaphors for language model governance, resulting in over two hundred metaphors: objects, places, processes, groups, and infrastructure that support reasoning about language model governance. What if a language model was like a community garden? Or a seed bank? Or the bathroom in a dive bar? We report on four themes: (1) the importance of consent, (2) how to define community boundaries, (3) ways to give contributor recognition, and (4) trade-offs in scale of language models. These metaphors point towards smaller, open models that encode group values. We discuss concrete ways to make community language models a reality. 


\end{abstract}



\begin{teaserfigure}
  \includegraphics[width=\textwidth]{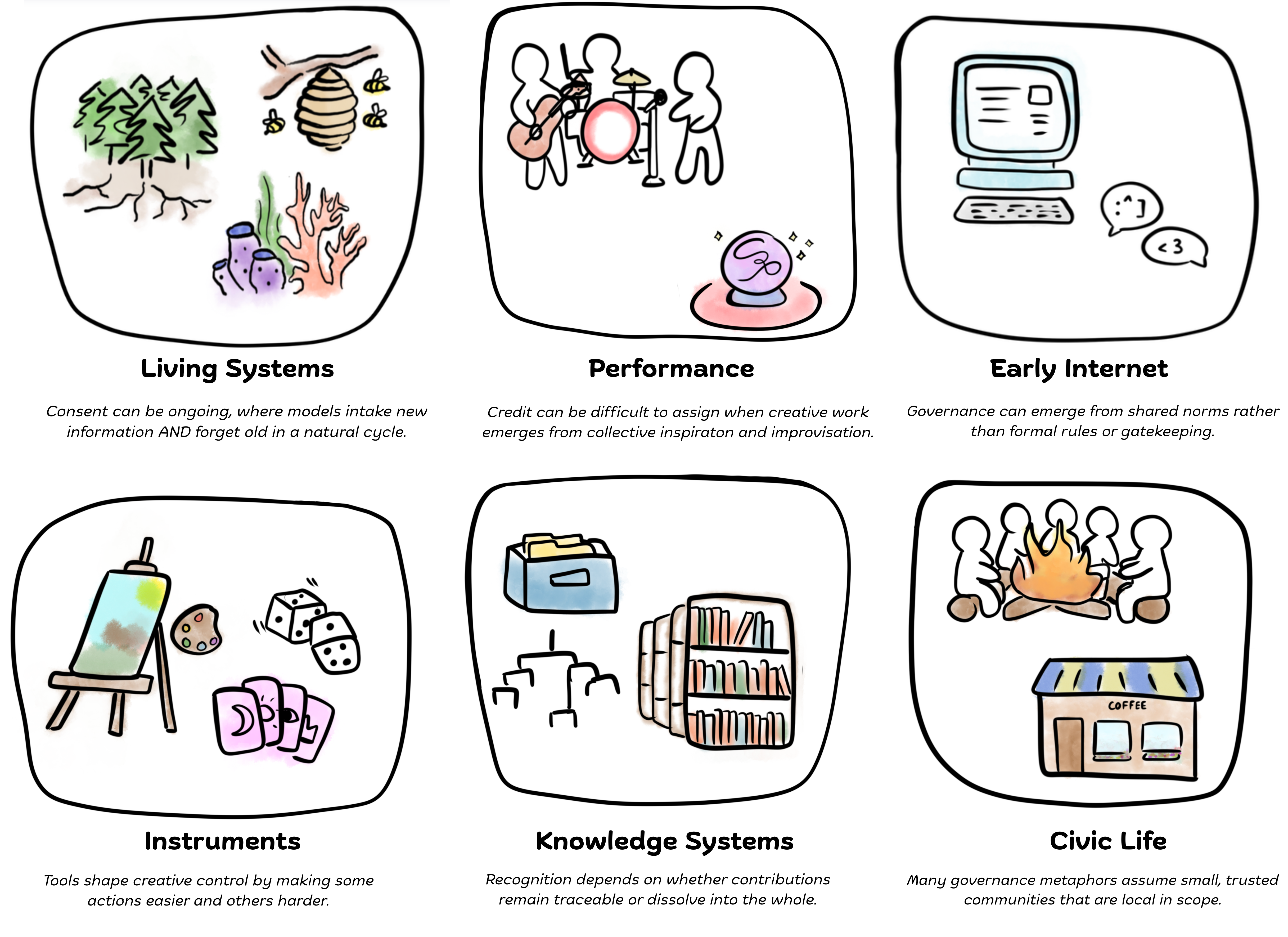}
  \caption{Six domains from which participants drew metaphors when reasoning about community governance of language models. Writers selected and extended metaphors from these domains to reason about values, relationships, trade-offs, and mechanisms of community language models.}
  \Description{A six-panel hand-drawn figure depicting metaphor domains used by participants: Living Systems, Performance, Early Internet, Instruments, Knowledge Systems, and Civic Life. Each panel shows simple illustrations (e.g., forests, a band, a computer, creative tools, archives, and a campfire or coffee shop) alongside short captions describing how each domain highlights different values in language model governance, including consent, credit, control, recognition, and scale.}
  \label{fig:teaser}
\end{teaserfigure}

\maketitle

\section{Introduction}

Creators of all stripes are alarmed by the development of generative AI models. For instance, large language models are often trained on massive datasets of scraped web text and books---often without authors' consent or even knowledge---and then marketed back to the public for generating the kind of writing authors produce professionally.
But not all creators are against the technology itself; instead they are troubled by the ways in which the technology is created, governed and used. Critiques of nonconsensual collection of training data \cite{authorsguildsurvey, asmelash2023thesebooksused}, the colonial or extractive nature of data use \cite{couldry2019DataColonialismRethinking}, and the negative impact on artist industries at large \cite{jiang2023AIArtIts} are critiques of a particular way of developing generative AI models, not (necessarily) of generative AI models as a concept.

But it is immensely difficult to separate the abstract idea of a technology from the way it is created and used. This is especially difficult for generative AI models where there are few examples that are created in line with values like respect for data contributors. While research has investigated the negative impacts of AI on artists \cite{jiang2023AIArtIts, lovato2024ForegroundingArtistOpinionsa} and some writers advocate for outright refusal or disengagement \cite{nplusone2025large}, less work has looked for positive futures in which creators can explore this technology on their own terms.

In this paper, we take a positive futuring stance and imagine how creative writers can claim this technology as their own. In particular, we focus on community governance, or how to build language models \textit{by and for }creative writers. Instead of advocating for better practices from existing model creators, we propose that communities can and should build this technology for themselves. \rv{We define ``community models'' as models built not only \textit{for} a community, but with meaningful participation from the community they are intended to serve---in their design and creation, not just as stakeholders or data contributors. This raises many questions about how community models may function, such as feasibility of model and community scale, how to collect and manage training data, whether such models can truly operate independently of commercial ones, and if they can ever feasibly be truly "large."}

To this end, we ran four workshops with over 100 creative writers to brainstorm new metaphors for language models that are created \textit{by and for} creative writers. Metaphors are powerful political tools that can reshape public discourse and corporate and governmental actions \cite{mio1997metaphor}. Writers were asked to (1) articulate how they would want a language model to make them feel, (2) brainstorm metaphors for alternative language model governance, and (3) analyze two metaphors in depth for how they would apply to language model governance. Our workshops and analysis respond to the following research questions:

\begin{itemize}
    \item How do writers envision positive or idealized community language model governance?
    \item What values do writers prioritize in community language model governance?
    \item How do these findings align with existing governance frameworks? 
\end{itemize}

Through our workshops and analysis, we report on the domains of metaphors that participants came up with (living systems, performance, early internet, civic life, knowledge systems, and instruments) and how these domains reflect the values of collective participation, care, and knowledge-sharing. When analyzing metaphors in depth, we found that participants came back to four main themes on how to manage community models: (1) the importance of \textbf{consent} as an ongoing, dynamic, and temporal process, (2) how to define implicit \textbf{boundaries} that could maintain control without explicit gatekeeping, (3) ways to give \textbf{recognition} to contributions to community models, and (4) trade-offs in \textbf{scale} that pushed participants towards desiring smaller models that could be run locally and preserve privacy.

These findings provide a nuanced understanding of the 3Cs framework for data contributors \cite{bota_moisin_3cs2017}, echoing some previous findings on the importance of consent and credit \cite{kyi2025GovernanceGenerativeAI} but differing from these findings on the importance of compensation, which was not prioritized by our participants in considering community models, and extending the analysis of the 3Cs framework to the context of community governance.

Based on our findings, we propose concrete strategies for how writers and researchers may work towards this goal. While this may seem idealized, we argue that recent technological advances show the potential of smaller models to better serve specific communities, and thus a viable path for creators to create their own generative models. We also discuss how our work does not necessarily invalidate advocacy for refusing or regulating commercial models trained on unconsensually collected data, and consider how community models may be served by concepts like data trusts \cite{delacroix2019BottomupDataTrusts} or data collectives~\cite{micheli2020EmergingModelsData}.

The contributions of this work are:

\begin{itemize}
    \item over 200 metaphors for AI governance \rv{generated by creative writers} that can be used for future design inquiries,
    \item four themes (consent, boundaries, recognition, and scale) that characterize writers' values for governance, and
    \item concrete suggestions for researchers and creative practitioners working to develop community models.
\end{itemize}


\section{Related Work}

\subsection{Writers' relationship with AI training data}
Algorithmic processes have long been used in writing practices, from tarot decks as storytelling devices, dice rolls deciding narrative elements, to computational rule-based systems to generate text. Recent approaches rely on training data rather than randomness or rules to create general purpose language models for a wide variety of tasks. Research has found that large language models can improve writing productivity and efficiency \cite{noy2023ExperimentalEvidenceProductivity}, and have been widely adopted in educational \cite{ravselj2025HigherEducationStudents} and professional contexts \cite{liang2025WidespreadAdoptionLarge}. But many creative writers object to the way these models are trained, in particular because many large language models have been trained on creative writing without the authors’ consent \cite{authorsguildsurvey, asmelash2023thesebooksused}, echoing broader concerns from artists \rv{around data extraction, economic loss, and devaluing human creative labor} \cite{jiang2023AIArtIts, lovato2024ForegroundingArtistOpinionsa, hong, HaveIBeenTrained, ScifiNPR}. 

Despite these concerns, some creative writers have \textit{chosen} to adopt these models in their practice. \citet{guo2025PenPromptHow}, studying creative writers who regularly use AI, find that writers are constantly negotiating their boundaries and making deliberate decisions on AI use according to their values. Notably, these writers still had concerns about the ways in which models were created and owned and were making pragmatic tradeoffs when using these tools.

\citet{gero2025CreativeWritersAttitudes} investigated how creative writers reason about the real or hypothetical use of their writing as training data. They found that creative writers have a nuanced understanding of LLMs and are more concerned with power imbalances in development and governance than the technology itself. \rv{The tensions become more visible given the growing ecosystem of commercial AI writing tools which have prompted backlash from writers around unclear training data provenance and concerns about replacing creative labor \cite{muldowney2025fanfiction, Viral, knibbs2023prosecraft}.} These findings are reflected by the proliferation of lawsuits against technology companies including OpenAI, Microsoft, Anthropic, and Google alleging violations of U.S. copyright law. As of early 2026, over 70 copyright infringement lawsuits have been filed against AI companies \cite{copyrightalliance2026}, with several reaching significant judicial decisions. In \textit{Bartz v. Anthropic} \cite{bartz_anthropic_order2025} courts found that using lawfully acquired books to train AI models was “transformative—spectacularly so” and constituted fair use, while simultaneously finding that Anthropic’s use of pirated works was not protected, leading to a \$1.5 billion settlement with authors of approximately 500,000 illegally downloaded works, the largest copyright settlement in U.S. history \cite{authorsguild2025anthropic}. In a related case, \textit{Kadrey v. Meta} \cite{kadrey_meta_order2025} found Meta’s use of copyrighted material constituted fair use, though the judge emphasized the narrowness of his holding. 

Across these contexts, consent is a top priority for writers, many of whom feel disempowered to stop the nonconsensual use of their work. Our work considers this context in which writers may be interested in large language models as a writing tool, but disturbed by the way in which they are developed and controlled.

\subsection{Data practices and governance in AI}

As machine learning moved towards requiring larger and larger datasets, researchers began to study “data work,” the practice of collecting, creating, curating, and managing data \cite{moller2020WhoDoesWork}. Research has highlighted the power imbalances in machine learning ecosystems where people working on data, as opposed to models or systems, typically have less power and prestige \cite{gero2023IncentiveGapData}. Concepts like data labor \cite{arrieta-ibarra2018ShouldWeTreat} and data leverage \cite{vincent2021DataLeverageFramework} have proposed ways to understand and even undo these power imbalances. 

A related body of work has emerged around governance as concerns around data practices have grown. Much of this work focuses on responsible or ethical AI frameworks \cite{lu2024ResponsibleAIPattern} or documenting existing governance structures \cite{batool2025AIGovernanceSystematic}. Such work considers how a collection of people, already in charge, are governing or should govern AI systems. Alternate governance structures have been proposed that give contributors more collective power over their data. Although many arose from concerns about power asymmetries in the platform economy (e.g., social media and gig work platforms), they could apply to generative AI data collection. For instance, data trusts \cite{delacroix2019BottomupDataTrusts} advocate that personal data could be held in trusts in which the trustee is legally obligated to advocate for the data contributors. \citet{micheli2020EmergingModelsData} identify four data governance practices—data trusts, data sharing pools, data cooperatives, and personal data sovereignty—that recognize contributors as stakeholders with governance rights and create intermediary structures to represent contributor interests. 

The 3Cs---consent, credit, and compensation---was originally developed by the Cultural Intellectual Property Rights Initiative for “sustainable, fair and equitable relationships with Traditional Knowledge and Traditional Cultural Expressions Custodians” \cite{bota_moisin_3cs2017}. This framework has been picked up in the AI community, sometimes appended with a 4th ‘C’: control \cite{sharma2025PRAC3PrivacyReputation, longpre2024position}. Most relevant to our work, \citet{kyi2025GovernanceGenerativeAI} investigate whether and how the 3Cs can serve as a useful framework for generative AI governance, interviewing a variety of creative workers (visual art, writing, and programming) and finding that they have nuanced understandings of the 3Cs, such as a desire for consent but acknowledging that existing power structures can make it difficult for them to provide meaningful consent. Due to its popularity in the AI community specifically, we draw on the 3Cs in analyzing our data.


In addition to theory and policy, several projects experimenting with community-governed models have begun to bring data contributors, model stewards, and users into a contiguous, collaborative ecosystem. One project worked with choirs to record a small-scale dataset explored attitudes on licensing and governing resulting models \cite{ding2024MyVoiceYour}. Bloom is a large-scale multilingual language model built collaboratively by hundreds of researchers worldwide, emphasizing open governance and transparency \cite{jernite2022DataGovernanceAge} and Latimer AI \cite{latimer2026usecases} aims to create AI systems that center Black and brown voices and histories. \rv{\citet{tseng} use participatory design with journalists to propose a language model that reflects the wants and challenges for AI in journalism. Such work demonstrates that members of particular communities can participate in generative AI development at the ``foundation" as well as ``subfloor" levels \cite{participatoryturn, tseng,suresh2024ParticipationAgeFoundation, vincent2021DataLeverageFramework}.}


These projects demonstrate growing interest in participatory AI development, though questions remain about how to scale community governance, manage diverse stakeholder interests, and sustain long-term collective stewardship. Our workshops contribute to this emerging landscape by surfacing the values, tensions, and governance models that creative writers themselves envision. 


\subsection{Metaphor as method for imagining alternatives} \label{RW_metaphor}
As Kenneth Burke writes in his “Four Master Tropes,” good metaphors allow us to see “something in terms of something else” \cite{burke1941four}. \citet{braunstein_warren_2021} expand on this Burkean impulse in their work on the stack metaphor in computing, writing: “As a rhetorical figure, metaphor shapes what can be thought. When it functions properly, we do not even notice the epistemic shifts that occur when one domain or scale substitutes for another.” Metaphors are both a rhetorical device and an epistemic instrument---they are doing simultaneous social and cognitive work. In this paper, we use metaphor generation to ask: what is epistemically prioritized when we imagine community data governance through frames like community gardens or anthills or garage bands?
 
Metaphors are recognized widely for their role in shaping cognition. In \textit{Metaphors We Live By}, \citet{lakoff1980metaphors} argue that our conceptual systems are fundamentally metaphorical, with abstract concepts like time systematically structured through mappings from more familiarly concrete, embodied, spatial experiences (e.g., looking forward, putting the past behind). Schön’s theorization of the “generative metaphor” moves from explanatory to imaginative---arguing that in addition to framing a way of looking at an idea or concept, metaphors can also be “a process by which new perspectives on the world come into existence” \cite{schon1979generative}. Phil Agre applies Schön’s “generative metaphor” in the context of computing and artificial intelligence \cite{agre1997computation}. Generative metaphors perform important selective work, determining the values, mechanisms, and relationships that are foregrounded and those that are marginalized in computing systems. 

HCI researchers have used generative metaphors as a method for reframing design problems \cite{lockton2019NewMetaphorsWorkshop, slipperyfish}, discussing and challenging dominant discourse around computational agency to open up design possibilities \cite{10.1145/3715336.3735714}, and productively framing ongoing tensions within the field via the “theory-practice gap” metaphor \cite{10.1145/3173574.3174194}. Taken together, these works—and many others---illustrate the power of generative metaphors as rhetorical, epistemic instruments that structure how researchers conceptualize problems, generate solutions, and understand disciplinary practices.

In our analysis, we draw on theories of generative metaphors and discourse communities as mechanisms for collective reasoning that participants went through to imagine alternative models for governance. 
These metaphors (or categories of metaphors) act as “boundary objects," or “objects that coordinate the perspectives of various communities of practice”~\cite{lee2007BoundaryNegotiatingArtifacts}. These objects can then be operationalized and made generative through conversation and collaboration. We also draw on the rhetorical concept of “discourse communities” to make sense of how participants came to assign meaning to these metaphors in the workshops. A discourse community is a “social group that communicates at least in part via written texts and shares common goals, values, and writing standards, a specialized vocabulary, and specialized genres” \cite{beaufort2007college}. Following Swales’ influential characterization, discourse communities develop through shared methods of communication, participatory practices, and collective expertise around common objectives \cite{swales1990genre}. In the context of our workshops, each instance of the workshop becomes its own discourse community, while each subgroup forms a smaller discourse community, developing their own vocabularies, values, and ways of reasoning about data rights and AI training practices. 



\section{Methods}
We conducted a series of four online workshops (around 25 participants each) over the first half of 2025 with creative writers interested in developing community-run language models “for and by” creative writers. Prior to these four workshops where we collected data, we first ran two pilot workshops to iterate on the workshop activities. The goals of the workshops were to surface writer perspectives on how they might collectively govern a language model for their own purposes, and to build a community of writers interested in exploring these alternatives. Since community building was just as important as addressing research questions, we minimized data collection to avoid hosting workshops that felt like we were ``studying'' the participants, especially given the sensitive context of AI in creative spaces \cite{muller2002}. The workshops led participants through exercises in generating and extending metaphors for model governance. \rv{Participation was voluntary and participants were not financially compensated for attending the workshops.} The data collected was analyzed using a modified thematic analysis approach. This study was approved by the relevant ethics review board.

\subsection{Participant Selection}
We aimed to gather participants who had a creative writing practice and were interested in more ethical versions of large language models that they might use in their practice. We recruited participants via the following channels: flyering around our home institutions, sharing a digital invite with relevant organizations,\footnote{e.g., The Writers Guild, the Authors Guild, and If, Then: Technology and Poetics working group, among others.} inviting people from our personal and professional networks, and snowball sampling (after each workshop, participants were encouraged to share the invite for future workshops). 
The workshop recruitment materials asked, “What Should a Language Model by Creative Writers, for Creative Writers Be? Join us for a Speculative Design Workshop.” To receive an invitation to the workshop, participants were asked to answer the following questions in a brief RSVP form: 

\begin{itemize}
    \item Do you write creatively? What do you like to write? What have you written before? What are you working on now?
    \item Do you use language models for creative writing? What do you like to use them for (if anything)? What do you dislike about them?
\end{itemize}

All respondents reported some kind of creative writing practice, and everyone was invited to participate. We did not collect explicit demographic information as it was not directly relevant to the goals of the workshop and our goal to minimize data collection. We received 131 RSVPs across the four workshops, although not every RSVP resulted in attendance. Approximately 25 people attended each workshop, plus 2 leaders (authors) and 3-5 facilitators (mix of authors and previous participants).

\begin{figure*}[!ht]
    \centering
    \begin{subfigure}[t]{0.24\linewidth}
        \centering
        \includegraphics[width=\linewidth]{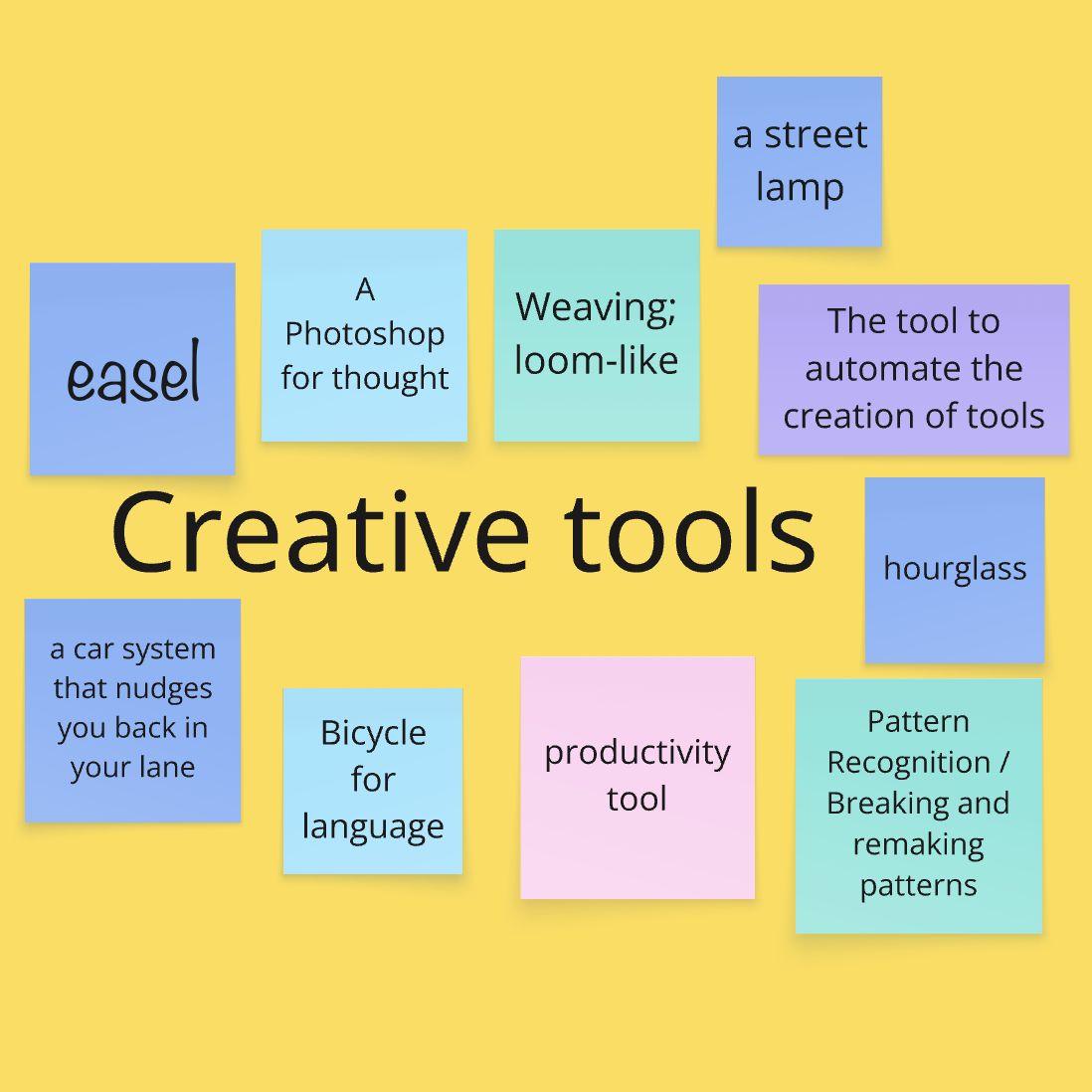}
        \caption{Cluster \textit{creative tools} with metaphors such as easel, bicycle for language, and 'weaving: loom like'.}
    \end{subfigure}\hfill
    \begin{subfigure}[t]{0.24\linewidth}
        \centering
        \includegraphics[width=\linewidth]{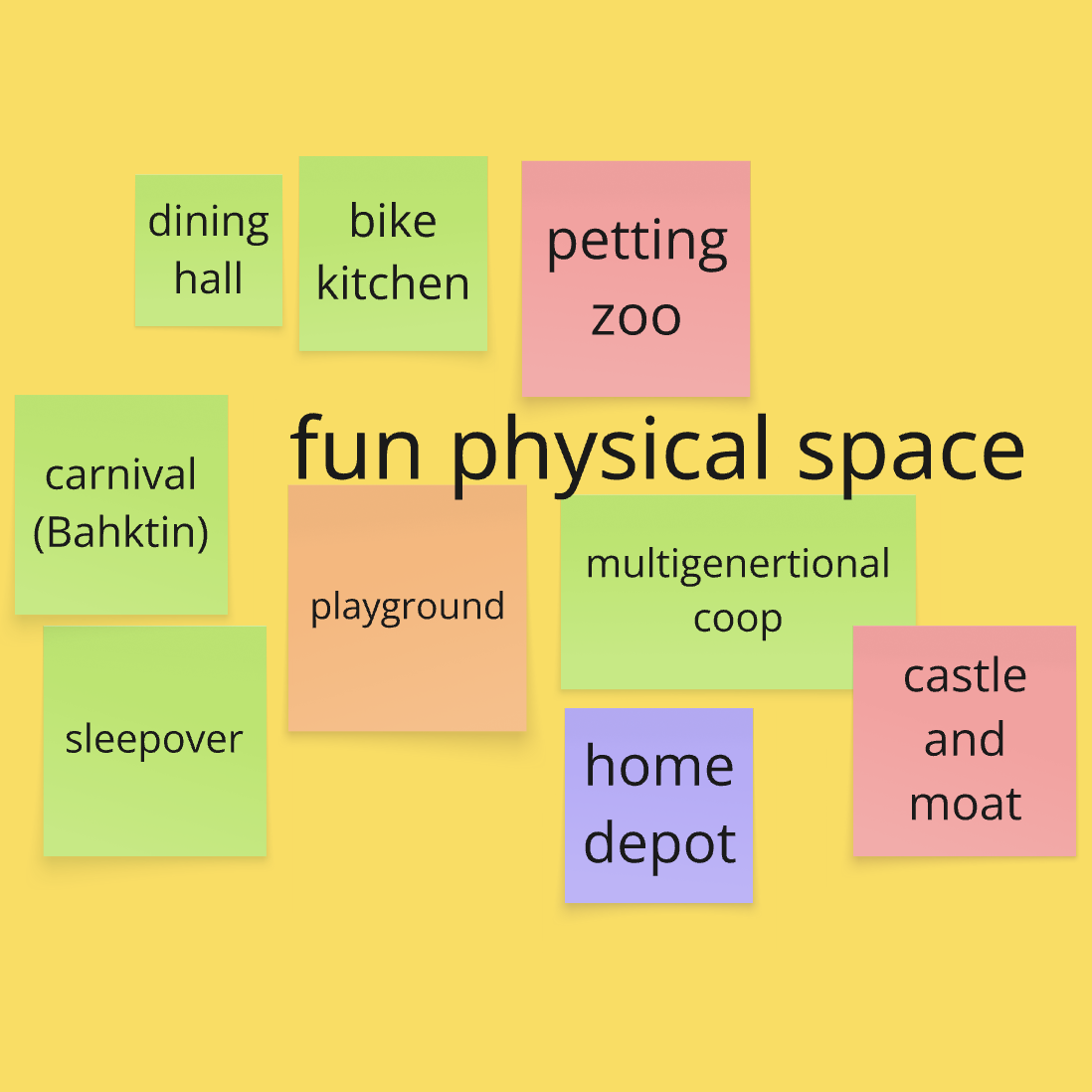}
        \caption{Cluster \textit{fun physical space} with metaphors such as dining hall, petting zoo, playground, and home depot.}
    \end{subfigure}\hfill
    \begin{subfigure}[t]{0.24\linewidth}
        \centering
        \includegraphics[width=\linewidth]{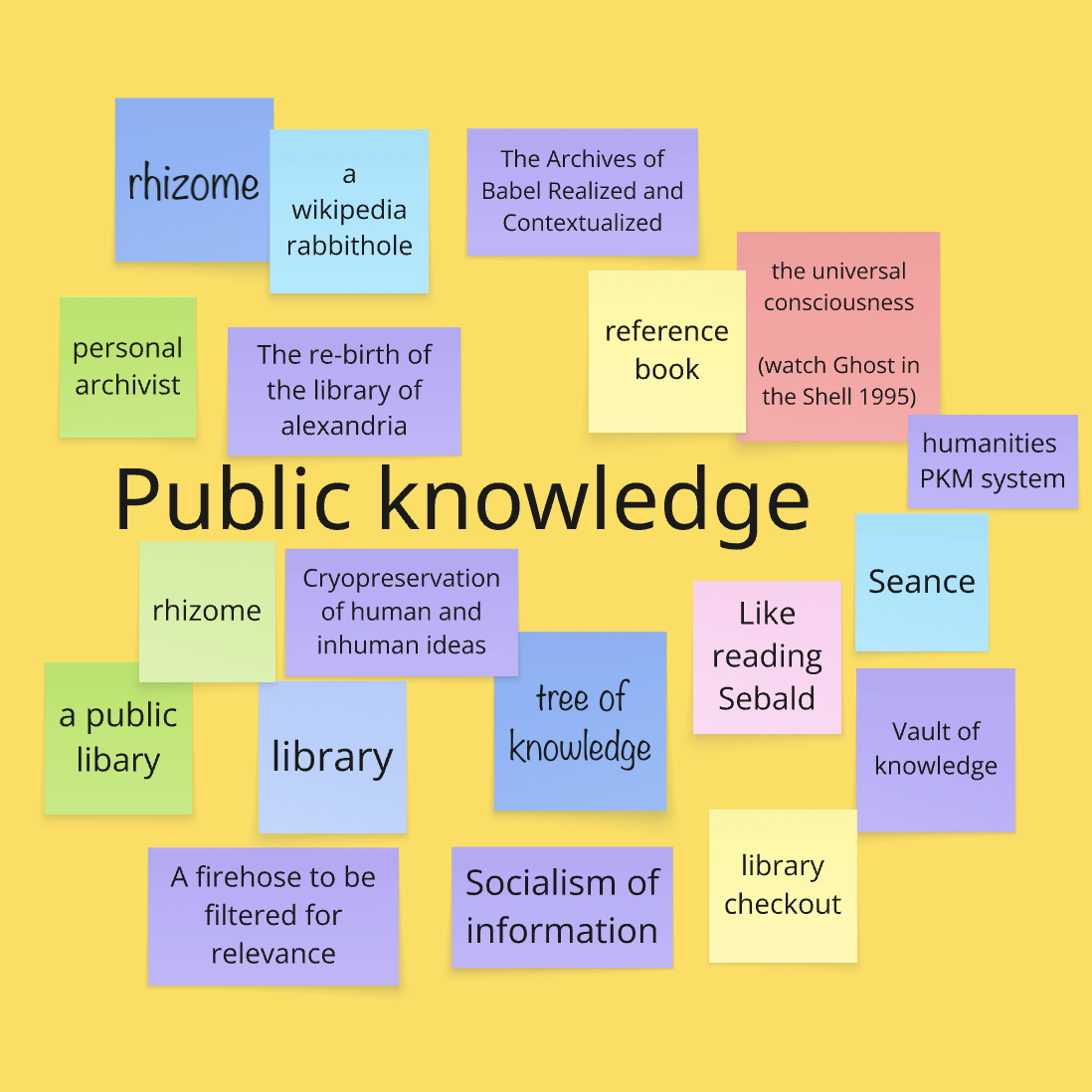}
        \caption{Cluster \textit{public knowledge} with metaphors such as tree of knowledge, library, reference book, and a wikipedia rabbit hole.}
    \end{subfigure}\hfill
    \begin{subfigure}[t]{0.24\linewidth}
        \centering
        \includegraphics[width=\linewidth]{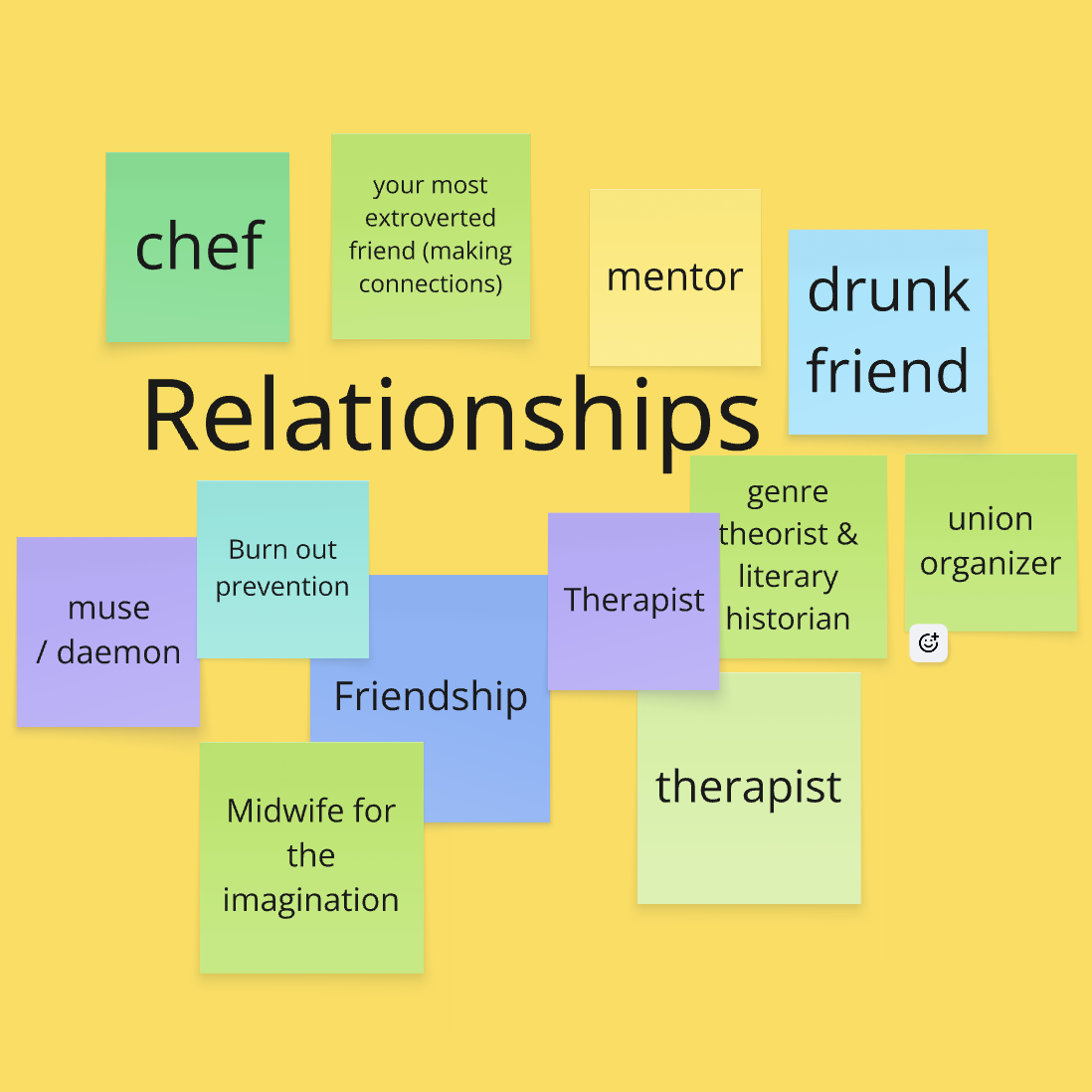}
        \caption{Cluster \textit{relationships} with metaphors such as mentor, chef, midwife for the imagination, and drunk friend.}
    \end{subfigure}
    \caption{Four clusters from the in-workshop clustering activity.}
    \label{fig:miroboardclusters}
\end{figure*}

Overall, our participants had a wide range of creative writing practices, with a little over half having prior experience with LLMs for creative writing. Participants' creative writing practices spanned a variety of genres (fiction, poetry, creative nonfiction, drama, speculative, academic, journaling) and levels of formality. Some participants held degrees in creative writing (undergraduate, MFA, PhD), and others taught writing in some capacity.
Some participants described themselves as “\textit{published writers}” whereas others noted they write “\textit{just for fun}” or “\textit{when the stars align.}” Many mentioned their ongoing projects, such as editing a novel manuscript or maintaining a Substack newsletter. A little over half of our participants reported having used LLMs in their writing practice in some capacity, citing tasks like brainstorming and world building or limiting use to proofreading, bringing up both the things they liked (efficiency, fluency) and disliked (robotic style). Those who had not used LLMs in their writing explained their reasons such as concerns about the training data and lack of creativity, while others were curious to try them in the future.

\subsection{Workshop Development}

Our workshops primarily used metaphor generation and analysis as a structuring activity. Similar to the ``new metaphors'' workshop method presented by \citet{lockton2019NewMetaphorsWorkshop}, we use metaphors to strategically reframe the problem of how to govern language models by considering other groups and processes that make collective decisions, and how writers might adopt those practices in this new context. Like in speculative design \cite{auger2013SpeculativeDesignCrafting}, the metaphors are intended to create a ‘perceptual bridge’ between participants’ perception of their world and the fictional element of a new way to create language models, while making space for critical reflection on values and ethics. Unlike speculative design, the metaphors are not intended to be potential designs in and of themselves, but rather ways to reshape how we think about language models and governance. \rv{Our workshop format followed prior methods for generating broad sets of metaphors, then choosing metaphors for group-based worksheet activities to elaborate on select metaphors \cite{lockton2019NewMetaphorsWorkshop}. We adapted this format through iteration and our own experiences participating in and organizing events for the creative writing community.}

\subsubsection{Workshop Iteration} Our workshop structure evolved through two pilot workshops, which were advertised in similar ways to the final four ``official'' workshops and also had about 25 participants each. Both pilot workshops used generating and analyzing metaphors as the structuring activity and were initially designed to be two parts of a set; participants were encouraged to attend both. The first focused on how writers would want to use language models in their practice and the second on how to govern them. We found that the ``usage'' workshop recovered familiar ideas about playful creativity support tools, so to maximize efficiency of participants' time (and logistical difficulty of attending two workshops) we rolled them into one, with a warm-up exercise about model usage and the main activity focusing on governance. Additionally, our initial pilot workshop on governance used a `worksheet,' designed around the 3Cs framework of consent, credit, and compensation, asking participants to think through the governance implications of the metaphors. We found that participants struggled to think directly about the 3Cs in this context and instead preferred to reason through the roles and relationships involved. We therefore redesigned the worksheet to focus on stakeholders, which led to richer participant engagement and discussion.

\subsection{Workshop Structure}
Each workshop lasted 90 minutes and had 1-2 leaders (authors) and 3-5 facilitators (authors and previous attendees). Previous attendees were invited to become facilitators for future workshops. Five participants ended up becoming facilitators and were given a short training session on their role during the workshop.\footnote{\rv{There was no screening process; everyone who wanted to be a facilitator after attending one of the workshops was welcomed. Some attended multiple workshops as facilitators. Facilitators were not financially compensated and were acknowledged during workshop sessions and in the workshop materials.}} A detailed description of the workshop procedures can be found in the Appendix; here we present a high-level summary.


\begin{figure*}[h]
    \centering
    \includegraphics[width=0.95\textwidth]{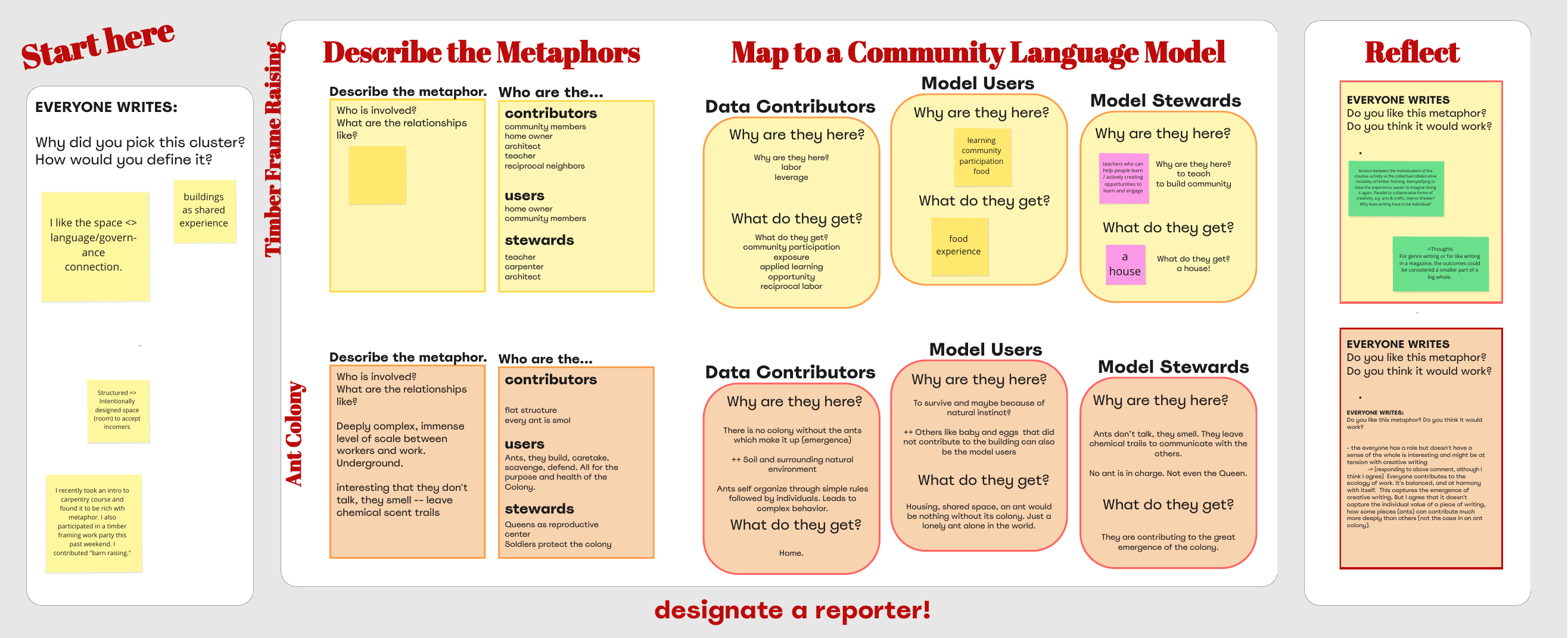}
    \caption{Worksheet provided to participant groups in the Miro board; this one is filled out by a group. The worksheet encouraged participants to talk through (and record) why they chose this cluster, how they define it, who the contributors, users, and stewards are, what they get out of the system, and reflect on what the participants thought worked or didn't work about the metaphor.}
    \Description{ Screenshot of a worksheet used in a collaborative digital whiteboard during a workshop, shown partially filled out by one group. The worksheet is divided into labeled sections that guide participants through the activity, including prompts to describe a chosen metaphor, identify contributors, users, and stewards, map these roles to a community language model, and reflect on whether the metaphor would work.}
    \label{fig:exampleworksheet}
\end{figure*}

\paragraph{Introduction and warm-up} Workshops began with the leaders introducing the goals of the workshop (to think through how to run a community language model), defining key terms (what is a language model, what we mean by metaphors for language model), setting the tone for the workshop, laying out the agenda and introducing the facilitators.\footnote{\rv{We note that participants were primed to think about community governance. This was because in the pilot workshops participants often veered into discussing how they might use a language model, rather than how they might \textit{create} a language model. Additionally, participants were primed to think about community governance, as opposed to corporate or national/regional governance.}} Participants then optionally introduced themselves in the chat. We then did a warm-up activity where participants described ``How you would want an interaction with a language model to make you feel?''

\rv{We introduced community models as a speculative alternative in which the communities a model is intended to serve would have greater involvement over all parts of the process of creating a model stemming from contributing their own writing for training, to the ultimate governance and use. We intentionally avoided presenting a fixed technical implementation (e.g., whether such models would be built from scratch, fine-tuned, or layered on existing systems) in order to preserve participants’ ability to define these systems themselves through metaphor. To get everyone on the same page, we introduced language models broadly as computer programs that generate text from large amounts of source text or training data and mentioned some popular commercial examples. We explicitly framed the activity as a space for reasoning together about governance that did not require technical expertise.}

\paragraph{Metaphor activities} We then moved to a Miro board---a collaborative digital whiteboard. Participants brainstormed metaphors for community-governed models as sticky notes on the whiteboard. After silent brainstorming, the facilitators, supported by the participants, would cluster the sticky-notes live during the workshop. \autoref{fig:miroboardclusters} shows a few clusters from the workshops. Then participants self-selected into breakout rooms based on the clusters and analyzed one to two metaphors from their cluster. We provided a worksheet in Miro to guide participants through questions about their specific metaphor in detail, focusing on three stakeholder groups: data contributors, model stewards, and model users. Each group had at least one facilitator to support and guide their discussions. We show an example worksheet, filled in by participants, in \autoref{fig:exampleworksheet}. 

\paragraph{Reporting back and closing.} Each group had a few minutes to report back about what they discussed in their group. The leaders would sum up interesting points that had emerged, then talk about the next workshops and other ways to get involved.

\subsection{Data analysis} 
We employed a qualitative analysis of all textual data that emerged from the workshops. Our data included answers to the warm-up questions, generated metaphors, clustering performed during the workshop, analysis done by subgroups on specific metaphors, and group reflections. 
We perform both inductive analysis, looking for emergent themes in the metaphors, such as the values they surface, and deductive analysis, looking for how the 3Cs \cite{bota_moisin_3cs2017} appeared in these metaphors. 

Since our analysis includes both participants’ analysis during the workshops and our own post-hoc analysis, we modify a typical thematic analysis approach \cite{braun2012ThematicAnalysis} to include the participant analysis as an initial coding process that is furthered by the authors of this paper. Our analysis additionally draws on theories about generative metaphors and discourse communities outlined in Section \ref{RW_metaphor}. \rv{All illustrations in the paper were hand-drawn by the first author.}


\begin{figure*}[!t]
    \centering
    \includegraphics[width=0.85\linewidth]{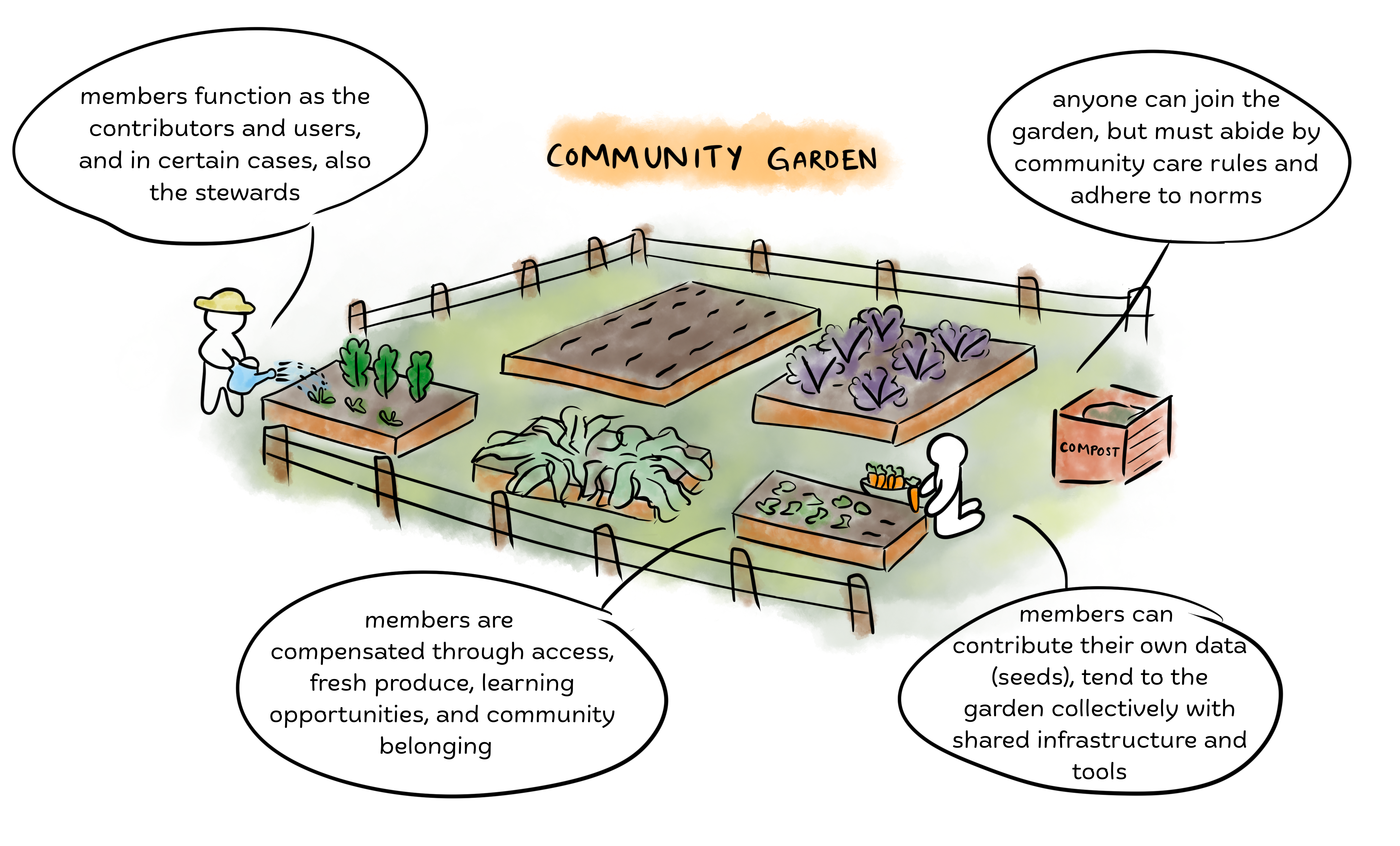}
    \caption{Expansion of the community garden metaphor \rv{from Group 3}, illustrating how participants mapped roles and governance mechanisms onto a shared language model. \rv{Text was paraphrased by authors for clarity.}}
    \Description{Hand-drawn illustration of a community garden used to explain governance concepts discussed by participants. The garden contains multiple raised beds with plants at different stages of growth, a compost bin, shared tools, and several people tending individual plots. Text callouts describe key aspects of the metaphor, including members acting as contributors, users, and sometimes stewards; participation being open but governed by shared norms of care; members contributing their own data while relying on shared infrastructure; and compensation taking the form of access, produce, learning, and community belonging.}
    \label{fig:garden}
\end{figure*}

\subsection{Positionality}
All authors of this paper have a creative writing practice, as well as experience in computing and ethics. Two of us are formally trained in Computer Science; one in English and Rhetoric. This research was conducted primarily in a U.S. context, although we had participants from North America, Europe, and Asia. We have the privilege of being affiliated with R1 academic institutions, which enabled us to put time towards this work. To the extent possible, we involved participants as equals in the research process, bringing them in as facilitators to increase the diversity of positions during the workshops.

We know and agree with many of the serious critiques of LLMs, both in the context of creative writing and otherwise, that often end with advocating refusal of or disengagement from the AI ecosystem. In our workshop opening we addressed the seriousness of those critiques, then argued that we saw a need for positive futuring as an important element of critique; we would also note that advocating for refusal was a common stance in many circles, and we wanted to provide space for other lines of advocacy. 


\section{Findings}

We report first on what the collection of generated metaphors, as a whole, tell us about participant values, as well as the domains and structures they represent. Then, we report on how participants analyzed specific metaphors and their implications for language model governance.

\subsection{Metaphor clusters and the values they represent}
\label{sectionfour}

Workshop participants generated 259 metaphors for model governance over the four workshops during the initial brainstorming session. \autoref{fig:miroboardclusters} shows clusters from one workshop, as clustered by the participants and facilitators during the workshop. Following the workshops, the authors conducted a second round of analysis, regrouping the 259 metaphors with the aim to identify patterns in metaphors across workshops, building on top of the initial groupings from the workshops rather than perfectly preserving them. Through this process, two dimensions emerged as we iteratively sorted metaphors and noticed recurring patterns in both what participants referenced (domain) and how they structured their thinking (structure). We use \textit{domain} to refer to the areas of life or objects of governance that participants drew from, such as early internet spaces or performance practices, which closely mirrored the quick categorizations made during the workshops. We use \textit{structure} to refer to the structural form and scale implied by the metaphor, for example whether the metaphor represents an object (e.g., quilt) or group (e.g., hiking club). 


\begin{table*}[!ht]
\centering
\renewcommand{\arraystretch}{1.25}
\begin{tabular}{m{1.2cm}m{3.5cm}m{8cm}}
\toprule
 & \textbf{Domain} & \textbf{Example metaphors from workshops} \\
\midrule
\centering\includegraphics[width=1cm]{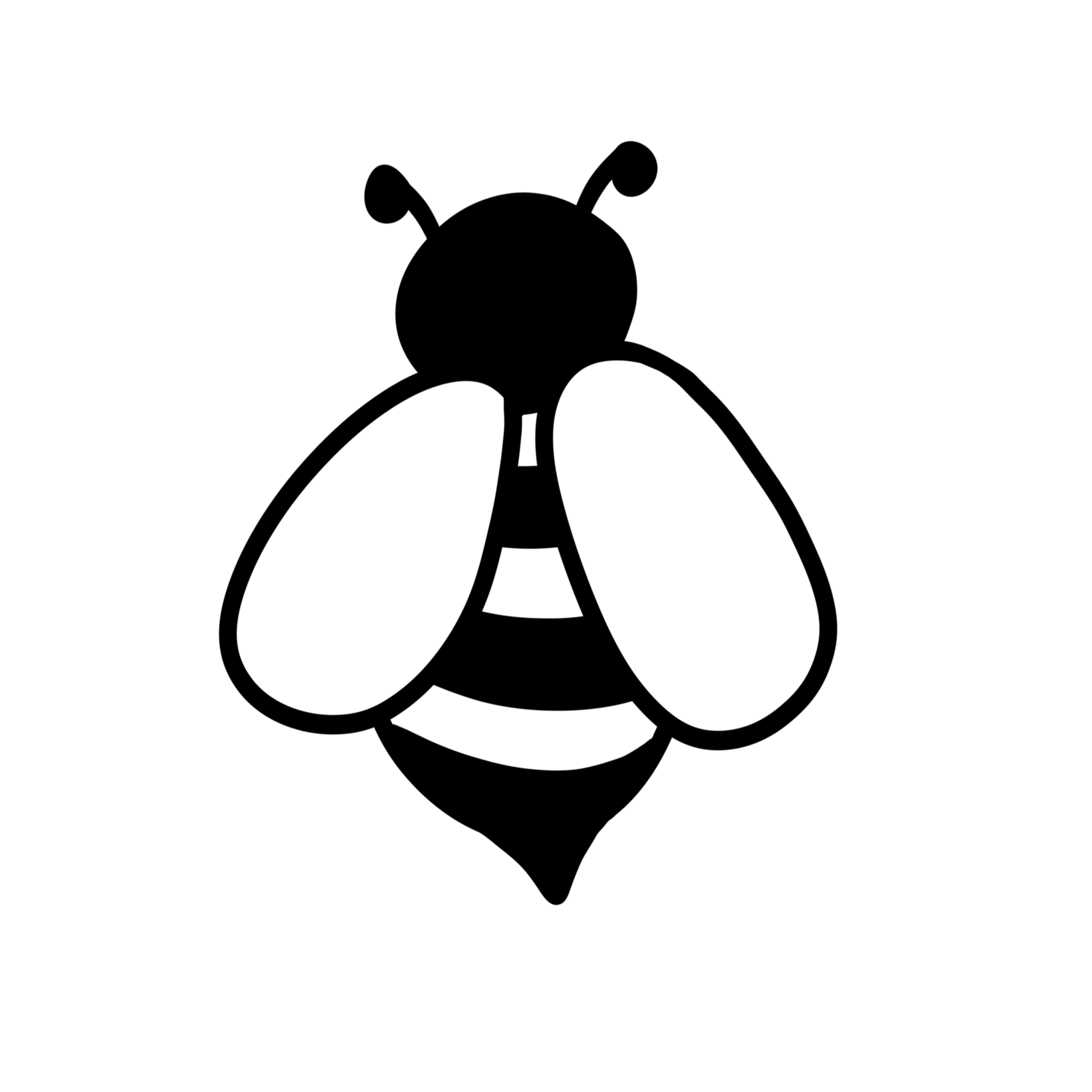} &
Living systems &
community garden, compost pile, SCOBY, meerkat colony, fungal network, beehive, ant colony, coral reef, aquarium \\
\midrule
\centering\includegraphics[width=1cm]{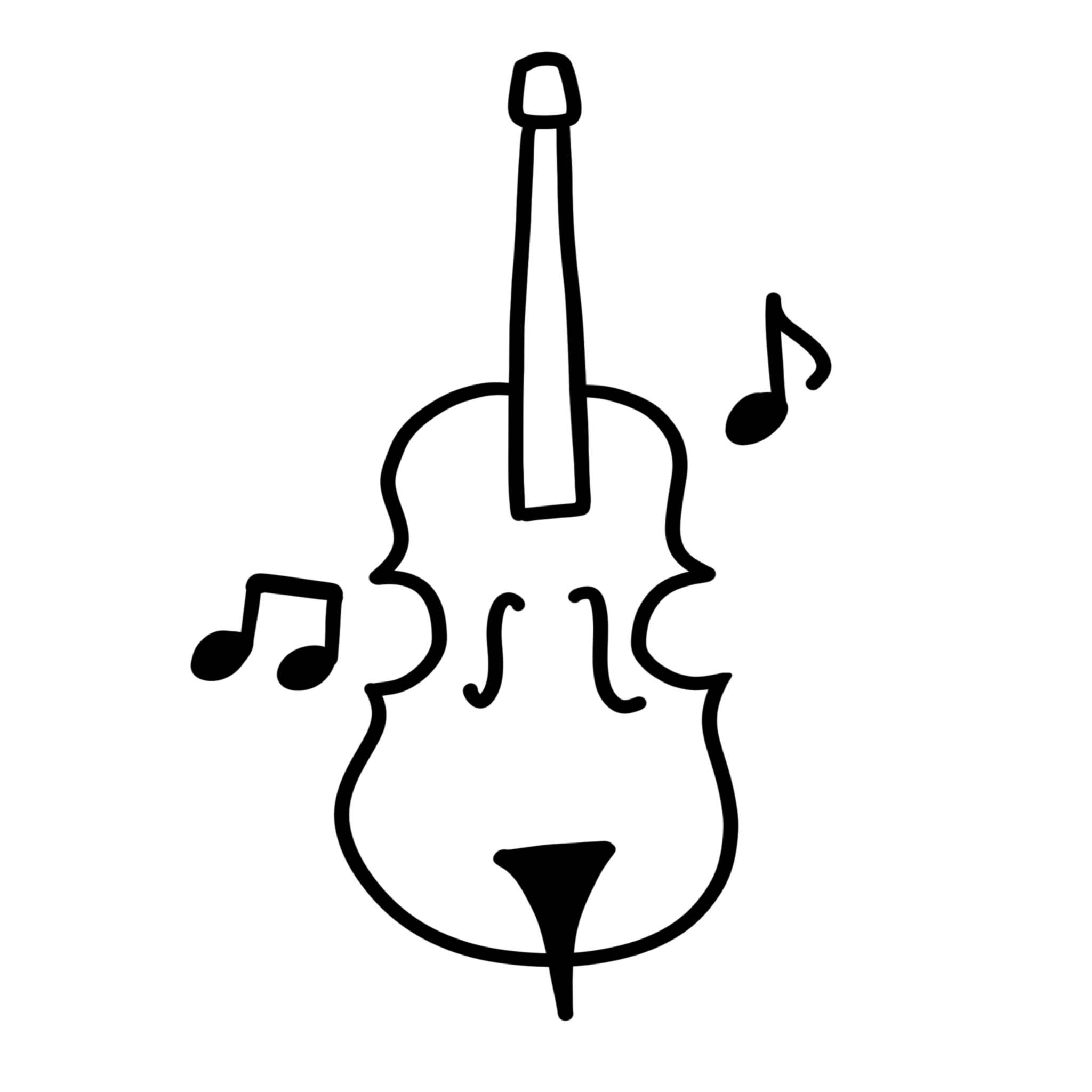} &
Performance &
band jam, choir, busking, seance, Burning Man \\
\midrule
\centering\includegraphics[width=1cm]{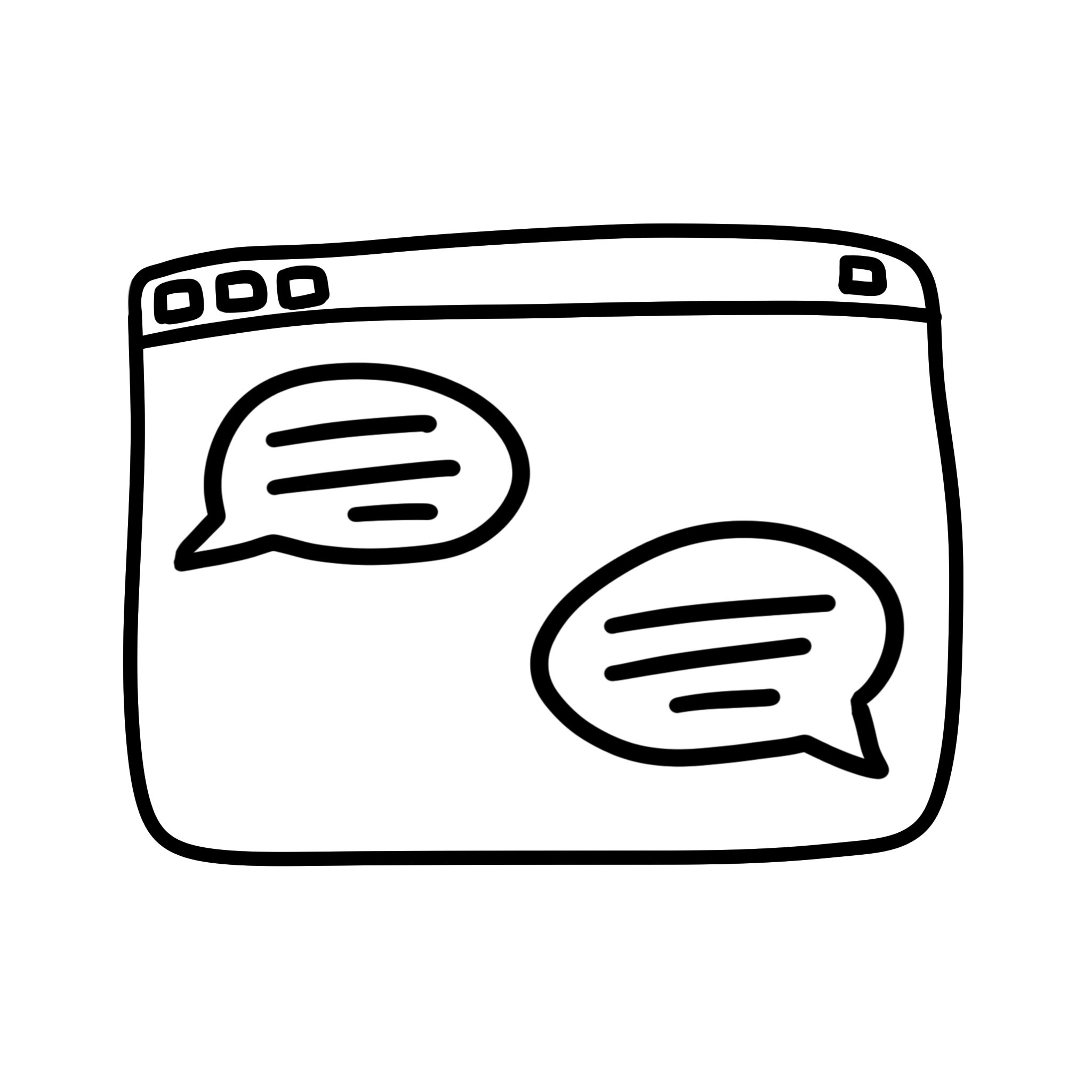} &
Early internet &
Geocities, Second Life, AO3, group text thread, Kaggle \\
\midrule
\centering\includegraphics[width=1cm]{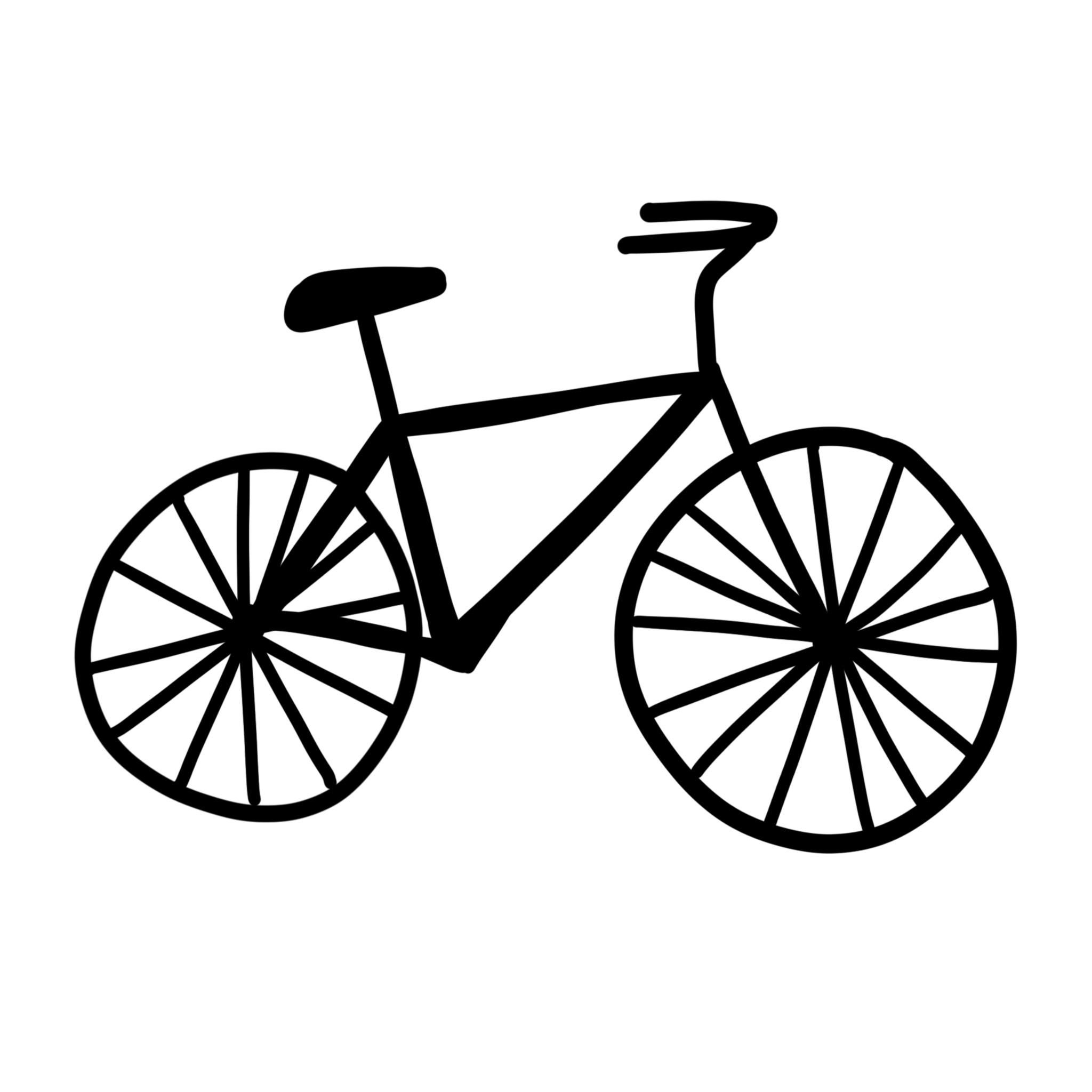} &
Civic life &
public library, coffee shop, seed bank, stoop swap, buy nothing groups, ham radio communities, solar microgrids, potluck, bbq, labor unions, hiking club, film club, priesthood, midwife, co-op, little free public library \\
\midrule
\centering\includegraphics[width=1cm]{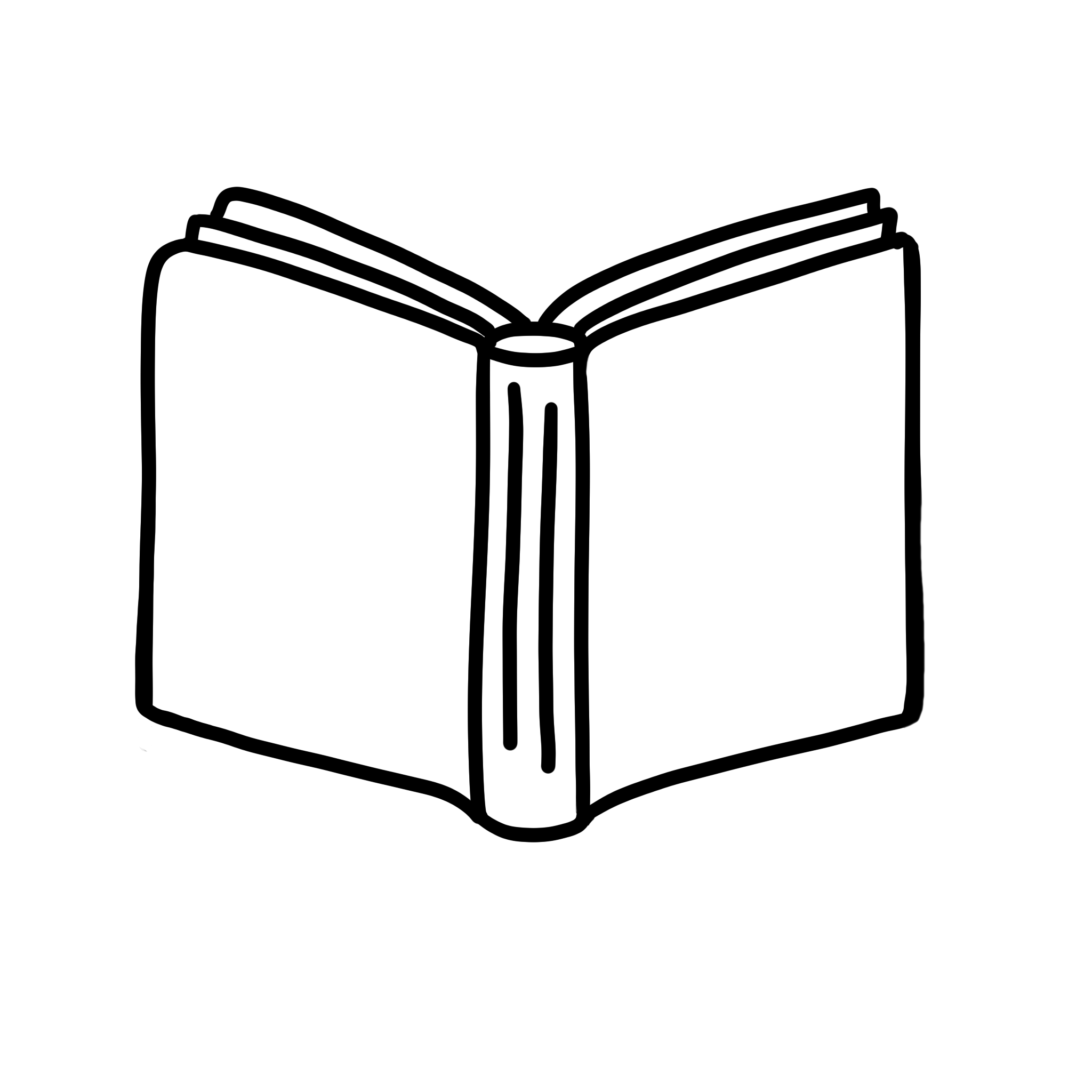} &
Knowledge systems &
community archives, family records, tree of knowledge, Wikipedia, library \\
\midrule
\centering\includegraphics[width=1cm]{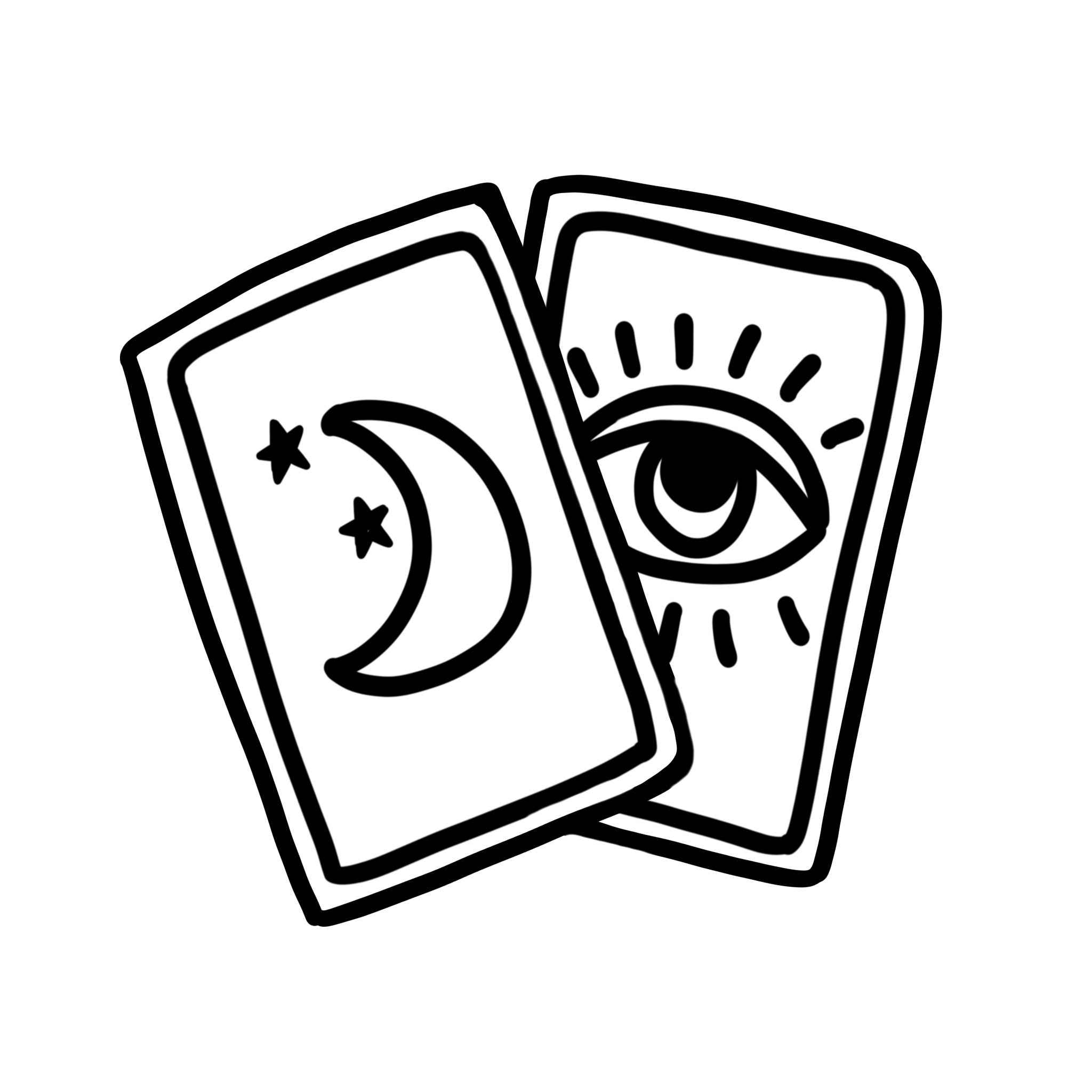} &
Instruments &
easel, Photoshop, AutoCAD, tarot cards, dice game \\
\bottomrule
\end{tabular}
\caption{Metaphors from all four workshops were clustered in six domains that represent the areas of life that participants drew from, largely characterized by collective participation, care, knowledge-sharing, and creative production.}
\label{tab:metaphor_domains}
\end{table*}

These dimensions serve to illustrate how participants initially approached thinking of metaphors for model governance through familiar experiences and desired quality or tone, while moving through different levels of abstraction. Domain and structure emerged organically in our analysis, which aligns with prior work that treats metaphors as selective and generative, deciding what values, mechanisms, and relationships become central versus peripheral in participants’ thinking about language model governance \cite{agre1997computation, schon1979generative}. 

The \textit{domains} (see \autoref{tab:metaphor_domains}) reveal the kinds of systems participants found most resonant or legible as models for governance. These metaphors were part of a brainstorming exercise, \rv{so no single metaphor necessarily represents participant values}; participants were also primed to consider community activities. That said, metaphors drew from domains characterized by care, knowledge-sharing, or creative production. Many metaphors eschewed traditional corporate or state structures, but some metaphors entailed commercial activities (coffee shop), commercial software (AutoCAD, Photoshop), or government-funded projects (public library), suggesting corporate and state structures can have elements of communal control. Notably, many metaphors were small in scale (band jam, film club, potluck) and prosocial, with many of them being playful and imaginative (beehive, seance, drunk friend).

The \textit{structures} (see \autoref{tab:structure_metaphors} in Appendix) reflect the angle from which participants approached a metaphor’s meaning. Objects suggest tools that can be wielded or modified; places imply spaces that require maintenance and access considerations; groups center collective decision-making and membership; processes emphasize ongoing evolution and participation; and infrastructure highlights questions of scale, interoperability, and long-term sustainability.
Most of these structural categories are not mutually exclusive; rather a single metaphor can span multiple levels of structure while highlighting different sources of meaning.

For example, a community garden can be understood as a physical place, but also the group of people that maintain and use it. Similarly, an object (loom) can also be framed as a process (weaving) or a group (weavers). Infrastructure operates as systems in the background that enable other activity, often maintained by groups. Together, the forms of the metaphors and their structural overlaps suggest that participants understood governance as a multi-structured mechanism rather than a single activity by a person or small group. In this view, governance is enacted through multiple interdependent layers---the model and its data, the people governing it, the processes of creating and maintaining the model, and the larger systems that sustain its existence---all of which interact to shape how the model operates within its broader environment.

\subsection{Metaphor analysis and their governance implications}

During the second portion of the workshop, participants extended select metaphors to reason about model governance; \autoref{tab:workshopclusters} shows the metaphors selected and discussed. A single metaphor did not map to one model of governance, but could be extended in a multitude of ways. For example, a community garden can involve individual plots or shared ones (Figure \ref{fig:garden}); a forest can be managed through active intervention or natural succession. 
Groups encountered these interpretive choices as important sites for reasoning in their expansion of metaphors, for example the plush chair metaphor group (M20) reflected: “\textit{There are some contradictions in the metaphors...when you're thinking about chair design, either the designers are iconoclastic and identifiable like Herman Miller or something completely anonymous like IKEA. We see that the metaphor can go either way.}” Each of these contradictions was an opportunity for groups to probe the implications of different governance priorities and decide which direction to continue in. 

\begin{table*}
\centering
\renewcommand{\arraystretch}{1.25}
\begin{tabular}{lp{6cm}p{7cm}}
\toprule
 & \textbf{Group} & \textbf{Metaphor/s}\\
\midrule
W1 & Performance (G1) & DJ! Musical Creation (M1); Ouija Board (M2) \\
 & Building (G2) & Timber Frame Raising (M3); Ant Colony (M4) \\
 & Gardening (G3) & Municipal Composting Program (M5); Community Vegetable Garden (M6) \\
\midrule
W2 & Place-based grassroots resources/swaps (G4) & Little Free Library (M7); Lightning Maps (M8) \\
 & Shared Physical Space (G5) & Campfire (M9); Church (M10) \\
 & Infrastructure (G6) & Microgrids (M11) \\
 & Archives (G7) & Community Archive (M12); Family Record (M13) \\
\midrule
W3 & Public Knowledge (G8) & Library (M14) \\
 & Creative Tools (G9) & Photoshop (M15); Loom (M16) \\
 & Relationships \& Learning Groups (G10) & Midwife (M17) \\
\midrule
W4 & Animal/plant collectives (G11) & Forest/fungal network (M18) \\
 & People collectives and personal internet (G12) & Monastic order (M19) \\
 & Shapes, performance, cooking (G13) & Plush chair (M20); Band/orchestra (M21) \\
 & Outdoor stewardship (G14) & Crags/trail system (M22) \\
\bottomrule
\end{tabular}
\caption{Workshop clusters and metaphors analyzed.}
\label{tab:workshopclusters}
\end{table*}

We report on four themes that emerged from our analysis. Some of these themes mapped well onto the 3Cs framework of consent, credit, and compensation, although participants were guided, via the worksheet, to talk about roles rather than the 3Cs framework. 

\subsubsection{\textbf{Theme 1: Consent} as ongoing, dynamic, and temporal}
Participants consistently emphasized the need for ethically sourced contributions and stated that consent was a pre-condition for community governance. How participants saw and defined consent differed, with many approaches proposed such as opt-ins, renew options, and revocation---what was consistent was the non-negotiability of intentional consent. Many metaphors used donations as a way to gather data, such as in municipal composting (M5) where contributors brought in their collected compost material. Other metaphors such as library (M14) located consent further upstream, through the act of publishing. For some, intentional consent entailed participants knowing intimately what their data was being used for and the actions their contributions were enabling, pointing at the purpose of the models, asking whether downstream decisions and qualities affect consent: “\textit{Do we need a clear sense of what is a good or bad purpose in order to make an ethical model?}” 

Consent was rarely framed as a one-time decision, but as something ongoing that data contributors would actively participate in, asked through questions of what happens when a community becomes less active, or when contributors no longer agree with their past decisions. Participants grappled with whether models should persist or fade alongside community activity. As one participant put it: 

\begin{quote}
    “\textit{When a Discord server or group chat stops being as participatory, do you want the LLM to live on as a record of that, as an archive? Or do you want it to only exist when people are actively participating? So we talked about the need for a kill switch...or other ways of naturally allowing the context to decline.}” 
\end{quote} 

Multiple groups landed on consent structures that changed over time and allowed for a natural built-in expiration to be part of the consent mechanism, where contributors would have to actively renew their consent either through explicit renewal or by the contribution of new data. Metaphors from the ``living system'' domain, such as a forest, fungi network and community gardens, prompted participants to imagine models that could change both with the intake of new information and the clearing of old information. G11 envisioned: “\textit{A forest ecosystem can cycle...the size and the limitation of the data should be set so that the amount of information that’s cycled through is constant...it learns to forget.}”

\subsubsection{\textbf{Theme 2: Boundaries} and other mechanisms for maintaining control}

Many groups were against formal contracts and explicit gatekeeping of a community model, exploring the tension between accessibility and the shared values and trust that make governance work. A group focused on shared physical places (G5) surfaced how physical locations can be gated, and rejected the idea of having a language model be ``gated'' with a key. Other groups also resisted restricted access, with one expressing the desire for models to be “\textit{absolutely free for as many people as possible},” pointing to resources like Reddit and Wikipedia and questioning the “\textit{extent to which you can really do community work}” with commercial models. 

However, participants worried about who gets to contribute, recognizing that contributors shape model values and character, suggesting some curation might be necessary. This would require access controls, at least on the model creation side. G3 likened discarding data that wasn’t serving the community space to keeping invasive species out of a pollinator garden, which would be carried out by community gardeners; similarly, a bar bathroom graffiti wall could be regulated by the owners or staff. At the same time, some groups ran into the limits of their metaphors since it is more difficult to know exactly how training data contributions map to language model characteristics or behavior. Such gatekeeping also didn't align with their values, because attempting to formalize requirements was “\textit{just a harder thing to enforce...once you're putting limits or giving form to what kinds of contributions count, you start putting walls up; instead of fencing in your garden to help curate it, you're then leaving people out}” (G4). 


Other metaphors such as churches, dance halls, and campfires described spaces with softer membership boundaries that relied on shared values, norms, and guidance from existing community members to guide access, rather than explicit ``gates'' or rules. One group expanded on churches (M10) and viewed them as welcoming but with strong norms: “\textit{It's open to people...but they also share the value system...At the same time, there is a structure, there are priests.}” In such metaphors, anyone can join but only those with similar values are likely to stay, and there are barriers to take on formal roles. 

\subsubsection{\textbf{Theme 3: Recognition} of presence in models (credit but not compensation)}

Most participants explicitly minimized the importance of monetary compensation as a motivator, but rather focused on credit, recognition, or access to the model as what they might be gaining from contributing to the model or to its governance. Some metaphors led participants to consider a notion of payment, particularly for use, similar to paying to see an orchestra performance~(M21) or having an engineering middleman following the business model of microgrids serving a community (M11). Many groups framed compensation through social connection and other experiential benefits rather than monetary payment, describing what contributors would get being ``\textit{creative spark, shared experiences, and socializing}'' (G5), ``\textit{something missing in their life...whimsy}'' (G3) or ``\textit{joy of playing}'' (G13). Other groups, while bringing up a concern that volunteer-led projects often fail to sustain themselves, prioritized thinking through how to assign credit or recognize contributors. Several groups noted that writers influence each other in entangled ways that resist clear attribution, similar to sampling in music (M21). One participant compared this to writer workshops: “\textit{You're inevitably influencing each other in the same way that an LLM output is made up of these latent influences...it’s difficult to ascribe credit.}” Rather than discrete contributions, influences blend together, raising questions about whether attribution is even desirable. As one participant asked, “\textit{being able to be traced back to a creator---is that a good thing or a bad thing? Would we want to be anonymous?}”

Instead of fine-grained credit, participants anchored on wanting recognition of “\textit{time and effort put into community creation},” similar to communities like Wikipedia and Reddit. Participants discussed credit not only outside the model (attribution for contributions) but also recognition within it---having a distinctive, recognizable presence in the model's outputs. Reflecting on the graffiti-in-a-bar-bathroom metaphor, one participant noted: “\textit{The mark on the wall signifies something that you don't normally see in a language model. You don't have that sense of presence, but I really love that idea of leaving a mark.}” Rather than explicit attribution, participants imagined recognition through felt influence.

\subsubsection{\textbf{Theme 4: Scale} and the desire for smaller language models}

Scale considerations through size of community, size of the model, and longevity of the model became key factors in how the metaphors played out and also a point of tension where they broke. Although participants were primed to imagine what governing a large language model might look like, the metaphors they chose and the ways in which they expanded them repeatedly pointed at the desire to have smaller, personalized, and intentionally limited systems. The archive group explicitly asked in relation to an archive's ability to preserve provenance \textit{and } uncertainty: ``\textit{thinking about smallness or in relationship to data...can we imagine a large language model that doesn't seem godlike}?'' (G7).

The desires that groups mentioned such as ongoing consent, membership that stems from shared values and trust, and meaningful recognition were often discussed through metaphors that assumed limited scale of community, i.e., local, tight knit communities. Questions such as “\textit{what happens when an insular dataset meant for two people grows into a massive LLM?}” surfaced, complicating how far the expansions of the chosen metaphors could be applied before behaviors changed. For example, the bar bathroom graffiti wall example brought up what would happen if the wall became a famous piece of art, raising the stakes and discouraging further free use for it. 

Participants repeatedly voiced wants for models that were local and personalized, which often went hand in hand. They recognized there would be a utility tradeoff with a smaller model, but many expressed that they didn’t need a big model, and many groups mentioned wanting these models to exist locally on their computer, not in the cloud. Writing can be incredibly personal, and participants wanted a way to be able to contribute and use models without feeling like they were giving up their privacy or control over their writing: “\textit{one of the big desires is to have models that are local first...especially if folks are going to put a corpus of their work or an individual work of theirs into a model.}” Some wanted to be able to use these models for their personal writing and journals, thus many existing tools have not been an option to them. In talking about how these models would be used (that is, why the participants might want to be contributors to a model,) participants consistently emphasized creativity over efficiency and even welcomed the idea of friction if it supported their creative experimentation, citing how tools can be molded by the people using them: “\textit{the more you sit in a chair, the more it becomes shaped to you}” (M20).

This automatically puts some limits on the size of a model, but in return can serve to address more specific needs, where one group stated that what was “\textit{important about that metaphor [microgrids] is that it’s hyperlocal and can really address the needs of smaller communities”} (G6). 


\section{Discussion}

\begin{quote}
    “\textit{This group of people's shared code of ethics or rules---that struck us as interesting, I think, in part because there's often a sense that language models are supposed to be these neutral artifacts, especially the big ones that are being produced by the big companies right now. And the idea of having them [the language models] be reflective of a relatively tight-knit group living according to a certain code that might be deliberately separating itself from the world in some way, that was interesting to us.}” (G12)
\end{quote}
 
The quote above demonstrates how writers envisioned language models that were not only communally governed, but reflective of shared values. While the metaphor activity spurred writers to think about consent, boundaries, recognition, and scale of language models (the \textit{mechanisms} required to train and manage a language model), they also surfaced values that at times were held in tension. Creative writing is often deeply individual, with participants wanting private, local models and wanting to be recognized for their mark on the model. On the other hand, most metaphors dealt in collaborative activities in which individual contributions are secondary to a larger goal. When discussing the ant colony metaphor, one participant explained, “\textit{that everyone has a role but doesn't have a sense of the whole is interesting and might be at tension with creative writing.}” 

Our findings confirm some aspects of prior work but depart from it in other areas. \citet{kyi2025GovernanceGenerativeAI} find that creative workers prioritize consent and compensation, while credit remains contextual (some thought their employer should get credit, others wanted to reduce their responsibility). We extend the 3Cs to a different context---community models rather than corporate ones---and similarly found consent was central, but participants were less interested in compensation and more in recognition through visible influence on the model. One notable difference in the community context was the overlapping stakeholder roles. Prior work, like \citet{kyi2025GovernanceGenerativeAI} and \citet{gero2025CreativeWritersAttitudes}, considered language model governance (by corporations or perhaps large research or government enterprises) where data contributors have almost no overlap with model creators or even necessarily model users. Our metaphors consistently pointed towards structures where the contributors \textit{are} the creators and the users. 

Even in metaphors that have structured levels of contributors, e.g., Wikipedia where there are users and then various levels of editors, there are \textit{paths} to formal roles for anyone interested. In contrast, it is difficult to imagine how one might move from a prolific data contributor to a model developer in corporate models, which could explain why our findings about, e.g., the importance of credit and compensation, differ from past work.  


Our work envisions community language models as a way for writers to create the kinds of language models they would like to use. \rv{However, we acknowledge that this does not ``solve'' any of the problems of current large language models (e.g., nonconsensual data collection \cite{authorsguildsurvey, asmelash2023thesebooksused}, extractive nature of data use \cite{couldry2019DataColonialismRethinking}, negative impacts on creative industries  \cite{jiang2023AIArtIts}). Writers consistently foregrounded consent, which likely requires divesting from many current large language models built on nonconsensually collected data.} If large language models were to adopt better data collection practices, for instance by licensing creative writing data via publishing houses or other arrangements, then perhaps some commercial models may become ethically acceptable to writers. \rv{In such cases, community models could be built on top of such models or become less necessary altogether. However, this would still leave major design decisions outside writers' control.} Writers in our workshops were excited about smaller models that reflected small-group values or styles, not purely because of their ethical status but also because of how they would behave differently from existing ones. 

\section{Limitations} 

Our participants were creative writers willing to engage with AI's ethical futures; writers who strongly oppose AI or prioritize commercial concerns might envision governance differently. The workshops emphasized positive futuring as a complementary resistance strategy to refusal or boycott.

Our U.S.-based recruitment and English-language workshops likely influenced the metaphors and values that emerged, which may differ in other cultural contexts. To minimize discomfort around this sensitive topic and avoid extractive research dynamics, we intentionally limited demographic data collection and did not track attendance across sessions, preventing analysis of how individual backgrounds shaped contributions.

Metaphors proved valuable for surfacing relationships and tensions, but have inherent limitations: as generative tools rather than prescriptive frameworks, they helped participants explore high-level directions without requiring them to resolve technical details like consent mechanisms or legal frameworks. The speculative nature of our workshops also meant participants weren't testing governance structures in practice; a significant constraint given that governance unfolds over time and reveals complexities only through lived experience.

\section{Future Work}

Our workshops were not only a research endeavor, but also a means of gauging interest and creating momentum in building community language models \textit{by and for} writers. The metaphors surfaced by writers were not intended to be treated as final governance solutions, but as starting points for writers to reason about different aspects of community model governance. We return to the themes from our findings---consent, boundaries, recognition, and scale---\rv{to outline future work through a sociotechnical lens that can be taken up by interested parties working together.}

\rv{\paragraph{Consent as ongoing, revocable, and temporal.} Writers imagined consent as ongoing rather than a one-time agreement, wanting meaningful ways to update consent with attention to identity (e.g. How would it feel to keep outdated versions of writing in a model?). Operationalizing this would require technical infrastructure for tracking contributions and regularly updating datasets, extending current research into machine unlearning and model editing \cite{liu2025rethinking, 10.5555/3737916.3741347, 10.1145/3733816.3760757}. It would also require interfaces that support ongoing consent over the life cycle of a model: ways of engaging with the effects of writer contributions and options for realistic withdrawal. Given the orientation towards smaller and more temporary models, future work should ask what new forms of consent become possible under these conditions? Possibilities include consent based on \textit{use cases}  (private journaling vs. public publishing) and community models that are designed to expire, deteriorate, or be forked or retrained as the community changes.} 


\rv{\paragraph{Boundaries as soft and based on community norms.} Participants gravitated toward boundaries that were softer than formal contracts, relying on clear community norms and values for standards of conduct and moderation. Future work should explore how communities can define membership processes, stewardship, and accountability without resorting to exclusion or gatekeeping. This could draw on existing governance practices from community-moderated forums~\cite{Fiesler_Jiang_McCann_Frye_Brubaker_2018}, projects such as Wikipedia \cite{forte2009DecentralizationWikipediaGovernance} and open-source software~\cite{omahony2007emergence}. Writing nonprofits and community writing organizations, e.g., the Authors Guild and the Association of Writers and Writing Programs, are especially well positioned to facilitate broader conversations, provide training, and help communities articulate community norms instead of asking individual groups of writers to create them from scratch.}

\rv{\paragraph{Recognition through felt presence.} Participants were less focused on monetary compensation and more intrigued by a different conception of recognition through felt presence and impact in the models' outputs. While technical work on training data attribution and influence can help explain how inputs affect model outputs, future work should ask how contributors can understand that their writing mattered. A community model might carry the history, lineage, genres, values, and voices of its contributors as a collective, rather than tracing each output to a specific writer. Because recognition is also a social practice, this could be complemented by norms and forms of credit that live outside the model itself, such as contributor lists, bios, statements, shared documentation, or community rituals of acknowledgment.}

\rv{\paragraph{Scale breaking assumptions about largeness.} 
Through their metaphors, participants repeatedly returned to models that function locally or at smaller scales to align with values of privacy and control. Future work should therefore ask what small models can do well for creative writing communities, rather than measuring them only against the assumptions of frontier-scale systems. Existing benchmarks may not serve these models well if they primarily measure broad capability rather than whether a model prompts creativity or reflects the voice or values of a particular community \cite{eldan2023tinystoriessmalllanguagemodels, wang2024weaverfoundationmodelscreative}.
Given that small communities likely mean small datasets, future work should also ask how much and what type of data will shift model properties. Work on fine-tuning LLMs with limited data could be extended to artistic settings \cite{szep2026FinetuningLargeLanguage, choshen2026babylmturns4goes}. Open-source and nonprofit models with more transparent data practices may provide useful foundations \cite{olmo2025olmo3}, but writers still need ways to understand tradeoffs between size and capabilities.}


\rv{\paragraph{Alternatives and tradeoffs} Writers also do not need to wait to start experimenting; small groups can already begin experimenting with datasets, continuing a longer history of writers building and training their own models \cite{sloan2016writing}. At the same time, we acknowledge meaningful participation requires some level of understanding of this technology in order to have agency over core system decisions. For community models, this might mean writers need a baseline technical and governance literacy to decide when to train, fine-tune, refuse, adapt, license, share, or abandon a model or dataset. To that end, HCI and NLP researchers can help build interfaces, evaluation methods, sensemaking tools, and technical infrastructure, but should avoid becoming permanent intermediaries between writers and the models they hope to govern. Prior work warns that participatory AI often relies on experts or practitioners as mediators between technical systems and affected communities, which can limit stakeholders’ direct agency even when projects are framed as participatory \cite{participatoryturn}.}

\rv{We also acknowledge that community models are not the answer to every problem that exists with large language models, and challenges and critiques remain. Issues with existing language models may still be present, such as the unintentional reproduction of biases, overfitting to a shared style, or the financial and environmental cost of training and maintenance. The metaphors generated in these workshops expand the imaginative possibility space of what language models could be, but speak less to when such models should be built and used. For some writers and writing communities, refusal, and litigation against unconsensual data use, may be the most fitting and powerful course of action. Future work in this space should look not just at how to build community models, but also in which situations it can be most beneficial, or when other avenues such as using open-source models, participating in collective licensing agreements,  and advocating for data trusts \cite{delacroix2019BottomupDataTrusts} or data collectives \cite{micheli2020EmergingModelsData} are more appropriate.}

\rv{We see these options as a constellation of actions, and writers may want to participate in one, multiple, or none. One writer may advocate for boycotting frontier LLMs while simultaneously joining a data trust (that is legally mandated to protect their interests) that licenses out their data for scientific research only and takes legal action against disallowed use cases. Another writer may use frontier models for marketing purposes and contribute to a community model that they use to voice a character in their novel. Many writers will likely not participate in any way – neither in refusal or engagement. Part of our goal here is to open up the space of possibilities, and move towards community models as an option.}

\section{Conclusion}

We began this paper by asking, \textit{How might we govern a language model run for and by creative writers?} Through a series of workshops with over one hundred creative writers, we brainstormed and analyzed metaphors that helped us understand how community language models might be possible. Such models will likely have ongoing consent procedures, have informal boundaries based on shared values and required contributions, allow contributors to see their mark on the models, and scale down the model size to allow for better community dynamics. Writers, NLP researchers, and HCI researchers can all contribute to this goal, which would empower writers, and indeed any creative practitioners, to engage with generative AI models in ways that align with their values.



\bibliographystyle{ACM-Reference-Format}
\bibliography{main}

@misc{eldan2023tinystoriessmalllanguagemodels,
      title={TinyStories: How Small Can Language Models Be and Still Speak Coherent English?}, 
      author={Ronen Eldan and Yuanzhi Li},
      year={2023},
      eprint={2305.07759},
      archivePrefix={arXiv},
      primaryClass={cs.CL},
      url={https://arxiv.org/abs/2305.07759}, 
}

@misc{wang2024weaverfoundationmodelscreative,
      title={Weaver: Foundation Models for Creative Writing}, 
      author={Tiannan Wang and Jiamin Chen and Qingrui Jia and Shuai Wang and Ruoyu Fang and Huilin Wang and Zhaowei Gao and Chunzhao Xie and Chuou Xu and Jihong Dai and Yibin Liu and Jialong Wu and Shengwei Ding and Long Li and Zhiwei Huang and Xinle Deng and Teng Yu and Gangan Ma and Han Xiao and Zixin Chen and Danjun Xiang and Yunxia Wang and Yuanyuan Zhu and Yi Xiao and Jing Wang and Yiru Wang and Siran Ding and Jiayang Huang and Jiayi Xu and Yilihamu Tayier and Zhenyu Hu and Yuan Gao and Chengfeng Zheng and Yueshu Ye and Yihang Li and Lei Wan and Xinyue Jiang and Yujie Wang and Siyu Cheng and Zhule Song and Xiangru Tang and Xiaohua Xu and Ningyu Zhang and Huajun Chen and Yuchen Eleanor Jiang and Wangchunshu Zhou},
      year={2023}, 
month={Jan.}}

@article{Fiesler_Jiang_McCann_Frye_Brubaker_2018, 
title={Reddit Rules! Characterizing an Ecosystem of Governance}, 
volume={12}, 
url={https://ojs.aaai.org/index.php/ICWSM/article/view/15033}, 
DOI={10.1609/icwsm.v12i1.15033}, 
number={1}, 
journal={Proceedings of the International AAAI Conference on Web and Social Media}, author={Fiesler, Casey and Jiang, Jialun and McCann, Joshua and Frye, Kyle and Brubaker, Jed}, 
year={2018}, 
month={Jun.} }

@article{omahony2007emergence,
  title={The Emergence of Governance in an Open Source Community},
  author={O'Mahony, Siobh{\'a}n and Ferraro, Fabrizio},
  journal={Academy of Management Journal},
  volume={50},
  number={5},
  pages={1079--1106},
  year={2007},
  url={http://www.jstor.org/stable/20159914}
}

@inproceedings{10.1145/3733816.3760757,
author = {Slate, Daniel D. and Chen, Chaoran and Yao, Yaxing and Li, Toby Jia-Jun},
title = {Iterative Contextual Consent: AI-enabled Data Privacy Contracts},
year = {2025},
isbn = {9798400719059},
publisher = {Association for Computing Machinery},
address = {New York, NY, USA},
url = {https://doi.org/10.1145/3733816.3760757},
doi = {10.1145/3733816.3760757},
booktitle = {Proceedings of the 2025 Workshop on Human-Centered AI Privacy and Security},
pages = {84–91},
numpages = {8},
location = {
},
series = {HAIPS '25}
}

@inproceedings{10.5555/3737916.3741347,
author = {Longpre, Shayne and Mahari, Robert and Lee, Ariel and Lund, Campbell and Oderinwale, Hamidah and Brannon, William and Saxena, Nayan and Obeng-Marnu, Naana and South, Tobin and Hunter, Cole and Klyman, Kevin and Klamm, Christopher and Schoelkopf, Hailey and Singh, Nikhil and Cherep, Manuel and Anis, Ahmad Mustafa and Dinh, An and Chitongo, Caroline and Yin, Da and Sileo, Damien and Mataciunas, Deividas and Misra, Diganta and Alghamdi, Emad and Shippole, Enrico and Zhang, Jianguo and Materzynska, Joanna and Qian, Kun and Tiwary, Kush and Miranda, Lester and Dey, Manan and Liang, Minnie and Hamdy, Mohammed and Muennighoff, Niklas and Ye, Seonghyeon and Kim, Seungone and Mohanty, Shrestha and Gupta, Vipul and Sharma, Vivek and Chien, Vu Minh and Zhou, Xuhui and Li, Yizhi and Xiong, Caiming and Villa, Luis and Biderman, Stella and Li, Hanlin and Ippolito, Daphne and Hooker, Sara and Kabbara, Jad and Pentland, Sandy and Data Provenance Initiative},
title = {Consent in crisis: the rapid decline of the AI data commons},
year = {2024},
isbn = {9798331314385},
publisher = {Curran Associates Inc.},
booktitle = {Proceedings of the 38th International Conference on Neural Information Processing Systems},
articleno = {3431},
numpages = {46},
location = {Vancouver, BC, Canada},
series = {NIPS '24}
}

@article{liu2025rethinking,
  title={Rethinking machine unlearning for large language models},
  author={Liu, Sijia and Yao, Yuanshun and Jia, Jinghan and Casper, Stephen and Baracaldo, Nathalie and Hase, Peter and Yao, Yuguang and Liu, Chris Yuhao and Xu, Xiaojun and Li, Hang and Varshney, Kush R. and Bansal, Mohit and Koyejo, Sanmi and Liu, Yang},
  journal={Nature Machine Intelligence},
  volume={7},
  pages={181--194},
  year={2025},
  doi={10.1038/s42256-025-00985-0},
  url={https://doi.org/10.1038/s42256-025-00985-0}
}

@misc{choshen2026babylmturns4goes,
      title={BabyLM Turns 4 and Goes Multilingual: Call for Papers for the 2026 BabyLM Workshop}, 
      author={Leshem Choshen and Ryan Cotterell and Mustafa Omer Gul and Jaap Jumelet and Tal Linzen and Aaron Mueller and Suchir Salhan and Raj Sanjay Shah and Alex Warstadt and Ethan Gotlieb Wilcox},
      year={2026},
      eprint={2602.20092},
      archivePrefix={arXiv},
      primaryClass={cs.CL},
      url={https://arxiv.org/abs/2602.20092}, 
}

@article{muller2002,
author = {Muller, Michael and Druin, Allison},
year = {2002},
month = {01},
pages = {},
title = {Participatory Design: The Third Space in HCI},
journal = {Handbook of HCI}
}

@inproceedings{participatoryturn,
author = {Delgado, Fernando and Yang, Stephen and Madaio, Michael and Yang, Qian},
title = {The Participatory Turn in AI Design: Theoretical Foundations and the Current State of Practice},
year = {2023},
isbn = {9798400703812},
publisher = {Association for Computing Machinery},
address = {New York, NY, USA},
url = {https://doi.org/10.1145/3617694.3623261},
doi = {10.1145/3617694.3623261},
booktitle = {Proceedings of the 3rd ACM Conference on Equity and Access in Algorithms, Mechanisms, and Optimization},
articleno = {37},
numpages = {23},
location = {Boston, MA, USA},
series = {EAAMO '23}
}

@inproceedings{tseng,
author = {Tseng, Emily and Young, Meg and Le Qu\'{e}r\'{e}, Marianne Aubin and Rinehart, Aimee and Suresh, Harini},
title = {"Ownership, Not Just Happy Talk": Co-Designing a Participatory Large Language Model for Journalism},
year = {2025},
isbn = {9798400714825},
publisher = {Association for Computing Machinery},
address = {New York, NY, USA},
url = {https://doi.org/10.1145/3715275.3732198},
doi = {10.1145/3715275.3732198},
booktitle = {Proceedings of the 2025 ACM Conference on Fairness, Accountability, and Transparency},
pages = {3119–3130},
numpages = {12},
location = {
},
series = {FAccT '25}
}

@misc{knibbs2023prosecraft,
  title={Why the Great AI Backlash Came for a Tiny Startup You’ve Probably Never Heard Of},
  url={https://www.wired.com/story/prosecraft-backlash-writers-ai/},
  journal={Wired},
  publisher={Condé Nast Publications},
  author={Knibbs, Kate},
  year={2023},
  month={Aug}
}

@misc{muldowney2025fanfiction,
  title={Fanfiction Writers Battle AI, One Scrape at a Time},
  url={https://www.theverge.com/ai-artificial-intelligence/688640/fanfiction-ai},
  journal={The Verge},
  publisher={Vox Media},
  author={Muldowney, Decca},
  year={2025},
  month={Jun}
}

@misc{Viral, title={Confessions of a Viral AI Writer}, url={https://www.wired.com/story/confessions-viral-ai-writer-chatgpt/}, journal={Wired}, publisher={Condé Nast Publications}, author={Vara, Vauhini}, year={2023}, month={Sept}}

@article{slipperyfish,
author = {Desai, Smit and Twidale, Michael},
title = {Metaphors in Voice User Interfaces: A Slippery Fish},
year = {2023},
issue_date = {December 2023},
publisher = {Association for Computing Machinery},
address = {New York, NY, USA},
volume = {30},
number = {6},
issn = {1073-0516},
url = {https://doi.org/10.1145/3609326},
doi = {10.1145/3609326},
journal = {ACM Trans. Comput.-Hum. Interact.},
month = sep,
articleno = {89},
numpages = {37}
}

@misc{ScifiNPR, title={Sci-Fi magazine stops submissions after flood of AI generated stories}, url={https://www.npr.org/2023/02/23/1159118948/sci-fi-magazine-stops-submissions-after-flood-of-ai-generated-stories}, journal={NPR}, publisher={NPR}, author={Vincent Acovino, Halimah Abdullah}, year={2023}, month={Feb}}

@misc{HaveIBeenTrained,
  author = {{Spawning AI}},
  title = {{Have I Been Trained?}},
  year = {2025},
  url = {https://haveibeentrained.com/},
  note = {Accessed: 2025-02-06}
}

@article{hong,
author = {Hong, Joo-Wha and Curran, Nathaniel Ming},
title = {Artificial Intelligence, Artists, and Art: Attitudes Toward Artwork Produced by Humans vs. Artificial Intelligence},
year = {2019},
issue_date = {April 2019},
publisher = {Association for Computing Machinery},
address = {New York, NY, USA},
volume = {15},
number = {2s},
issn = {1551-6857},
url = {https://doi-org.offcampus.lib.washington.edu/10.1145/3326337},
doi = {10.1145/3326337},
journal = {ACM Trans. Multimedia Comput. Commun. Appl.},
month = {jul},
articleno = {58},
numpages = {16}
}

@article{arrieta-ibarra2018ShouldWeTreat,
  title = {Should {{We Treat Data}} as {{Labor}}? {{Moving Beyond}} “{{Free}}”},
  shorttitle = {Should {{We Treat Data}} as {{Labor}}?},
  author = {Arrieta-Ibarra, Imanol and Goff, Leonard and Jiménez-Hernández, Diego and Lanier, Jaron and Weyl, E. Glen},
  date = {2018-05-01},
  journaltitle = {AEA Papers and Proceedings},
  shortjournal = {AEA Papers and Proceedings},
  volume = {108},
  pages = {38--42},
  issn = {2574-0768, 2574-0776},
  doi = {10.1257/pandp.20181003},
  url = {https://pubs.aeaweb.org/doi/10.1257/pandp.20181003},
  urldate = {2026-01-26},
  langid = {english}
}

@article{batool2025AIGovernanceSystematic,
  title = {{{AI}} Governance: A Systematic Literature Review},
  shorttitle = {{{AI}} Governance},
  author = {Batool, Amna and Zowghi, Didar and Bano, Muneera},
  date = {2025-06},
  journaltitle = {AI and Ethics},
  shortjournal = {AI Ethics},
  volume = {5},
  number = {3},
  pages = {3265--3279},
  issn = {2730-5953, 2730-5961},
  doi = {10.1007/s43681-024-00653-w},
  url = {https://link.springer.com/10.1007/s43681-024-00653-w},
  urldate = {2026-01-26},
  langid = {english}
}

@article{delacroix2019BottomupDataTrusts,
  title = {Bottom-up Data {{Trusts}}: Disturbing the ‘One Size Fits All’ Approach to Data Governance},
  shorttitle = {Bottom-up Data {{Trusts}}},
  author = {Delacroix, Sylvie and Lawrence, Neil D},
  date = {2019-10-01},
  journaltitle = {International Data Privacy Law},
  pages = {ipz014},
  issn = {2044-3994, 2044-4001},
  doi = {10.1093/idpl/ipz014},
  url = {https://academic.oup.com/idpl/advance-article/doi/10.1093/idpl/ipz014/5579842},
  urldate = {2026-01-26},
  langid = {english}
}

@inproceedings{gero2025CreativeWritersAttitudes,
  title = {Creative {{Writers}}' {{Attitudes}} on {{Writing}} as {{Training Data}} for {{Large Language Models}}},
  booktitle = {Proceedings of the 2025 {{CHI Conference}} on {{Human Factors}} in {{Computing Systems}}},
  author = {Gero, Katy Ilonka and Desai, Meera and Schnitzler, Carly and Eom, Nayun and Cushman, Jack and Glassman, Elena L.},
  date = {2025-04-25},
  series = {{{CHI}} '25},
  pages = {1--16},
  publisher = {Association for Computing Machinery},
  location = {New York, NY, USA},
  doi = {10.1145/3706598.3713287},
  url = {https://doi.org/10.1145/3706598.3713287},
  urldate = {2025-07-22},
  isbn = {979-8-4007-1394-1}
}

@inproceedings{guo2025PenPromptHow,
  title = {From {{Pen}} to {{Prompt}}: {{How Creative Writers Integrate AI}} into Their {{Writing Practice}}},
  shorttitle = {From {{Pen}} to {{Prompt}}},
  booktitle = {Proceedings of the 2025 {{Conference}} on {{Creativity}} and {{Cognition}}},
  author = {Guo, Alicia and Sathyanarayanan, Shreya and Wang, Leijie and Heer, Jeffrey and Zhang, Amy X.},
  date = {2025-06-23},
  pages = {527--545},
  publisher = {ACM},
  location = {Virtual United Kingdom},
  doi = {10.1145/3698061.3726910},
  url = {https://dl.acm.org/doi/10.1145/3698061.3726910},
  urldate = {2026-02-03},
  eventtitle = {C\&{{C}} '25: {{Creativity}} and {{Cognition}}},
  isbn = {979-8-4007-1289-0},
  langid = {english}
}

@inproceedings{jiang2023AIArtIts,
  title = {{{AI Art}} and Its {{Impact}} on {{Artists}}},
  booktitle = {Proceedings of the 2023 {{AAAI}}/{{ACM Conference}} on {{AI}}, {{Ethics}}, and {{Society}}},
  author = {Jiang, Harry H. and Brown, Lauren and Cheng, Jessica and Khan, Mehtab and Gupta, Abhishek and Workman, Deja and Hanna, Alex and Flowers, Johnathan and Gebru, Timnit},
  date = {2023-08-08},
  pages = {363--374},
  publisher = {ACM},
  location = {Montr\textbackslash '\{e\}al QC Canada},
  doi = {10.1145/3600211.3604681},
  url = {https://dl.acm.org/doi/10.1145/3600211.3604681},
  urldate = {2025-02-02},
  eventtitle = {{{AIES}} '23: {{AAAI}}/{{ACM Conference}} on {{AI}}, {{Ethics}}, and {{Society}}},
  isbn = {979-8-4007-0231-0},
  langid = {english}
}

@article{lovato2024ForegroundingArtistOpinionsa,
  title = {Foregrounding {{Artist Opinions}}: {{A Survey Study}} on {{Transparency}}, {{Ownership}}, and {{Fairness}} in {{AI Generative Art}}},
  shorttitle = {Foregrounding {{Artist Opinions}}},
  author = {Lovato, Juniper and Zimmerman, Julia Witte and Smith, Isabelle and Dodds, Peter and Karson, Jennifer L.},
  date = {2024-10-16},
  journaltitle = {Proceedings of the AAAI/ACM Conference on AI, Ethics, and Society},
  shortjournal = {AIES},
  volume = {7},
  pages = {905--916},
  issn = {3065-8365},
  doi = {10.1609/aies.v7i1.31691},
  url = {https://ojs.aaai.org/index.php/AIES/article/view/31691},
  urldate = {2026-02-03}
}

@article{lu2024ResponsibleAIPattern,
  title = {Responsible {{AI Pattern Catalogue}}: {{A Collection}} of {{Best Practices}} for {{AI Governance}} and {{Engineering}}},
  shorttitle = {Responsible {{AI Pattern Catalogue}}},
  author = {Lu, Qinghua and Zhu, Liming and Xu, Xiwei and Whittle, Jon and Zowghi, Didar and Jacquet, Aurelie},
  date = {2024-07-31},
  journaltitle = {ACM Computing Surveys},
  shortjournal = {ACM Comput. Surv.},
  volume = {56},
  number = {7},
  pages = {1--35},
  issn = {0360-0300, 1557-7341},
  doi = {10.1145/3626234},
  url = {https://dl.acm.org/doi/10.1145/3626234},
  urldate = {2026-01-26},
  langid = {english}
}

@article{micheli2020EmergingModelsData,
  title = {Emerging Models of Data Governance in the Age of Datafication},
  author = {Micheli, Marina and Ponti, Marisa and Craglia, Max and Berti Suman, Anna},
  date = {2020-07},
  journaltitle = {Big Data \& Society},
  shortjournal = {Big Data \& Society},
  volume = {7},
  number = {2},
  pages = {2053951720948087},
  issn = {2053-9517, 2053-9517},
  doi = {10.1177/2053951720948087},
  url = {https://journals.sagepub.com/doi/10.1177/2053951720948087},
  urldate = {2026-01-26},
  langid = {english}
}

@article{moller2020WhoDoesWork,
  title = {Who Does the Work of Data?},
  author = {Møller, Naja Holten and Bossen, Claus and Pine, Kathleen H. and Nielsen, Trine Rask and Neff, Gina},
  date = {2020-04-17},
  journaltitle = {Interactions},
  shortjournal = {interactions},
  volume = {27},
  number = {3},
  pages = {52--55},
  issn = {1072-5520, 1558-3449},
  doi = {10.1145/3386389},
  url = {https://dl.acm.org/doi/10.1145/3386389},
  urldate = {2026-01-26},
  langid = {english}
}

@inproceedings{vincent2021DataLeverageFramework,
  title = {Data {{Leverage}}: {{A Framework}} for {{Empowering}} the {{Public}} in Its {{Relationship}} with {{Technology Companies}}},
  shorttitle = {Data {{Leverage}}},
  booktitle = {Proceedings of the 2021 {{ACM Conference}} on {{Fairness}}, {{Accountability}}, and {{Transparency}}},
  author = {Vincent, Nicholas and Li, Hanlin and Tilly, Nicole and Chancellor, Stevie and Hecht, Brent},
  date = {2021-03-03},
  pages = {215--227},
  publisher = {ACM},
  location = {Virtual Event Canada},
  doi = {10.1145/3442188.3445885},
  url = {https://dl.acm.org/doi/10.1145/3442188.3445885},
  urldate = {2026-01-26},
  eventtitle = {{{FAccT}} '21: 2021 {{ACM Conference}} on {{Fairness}}, {{Accountability}}, and {{Transparency}}},
  isbn = {978-1-4503-8309-7},
  langid = {english}
}

@article{noy2023ExperimentalEvidenceProductivity,
  title = {Experimental Evidence on the Productivity Effects of Generative Artificial Intelligence},
  author = {Noy, Shakked and Zhang, Whitney},
  year = 2023,
  month = jul,
  journal = {Science},
  volume = {381},
  number = {6654},
  pages = {187--192},
  publisher = {American Association for the Advancement of Science (AAAS)},
  issn = {0036-8075, 1095-9203},
  doi = {10.1126/science.adh2586},
  urldate = {2025-08-07},
  langid = {english}
}

@article{gero2023IncentiveGapData,
  title = {The Incentive Gap in Data Work in the Era of Large Models},
  author = {Gero, Katy Ilonka and Das, Payel and Dognin, Pierre and Padhi, Inkit and Sattigeri, Prasanna and Varshney, Kush R.},
  year = 2023,
  month = jun,
  journal = {Nature Machine Intelligence},
  volume = {5},
  number = {6},
  pages = {565--567},
  issn = {2522-5839},
  doi = {10.1038/s42256-023-00673-x},
  urldate = {2025-02-02},
  copyright = {All rights reserved},
  langid = {english}
}

@article{forte2009DecentralizationWikipediaGovernance,
  title = {Decentralization in {{Wikipedia Governance}}},
  author = {Forte, Andrea and Larco, Vanesa and Bruckman, Amy},
  year = 2009,
  month = jul,
  journal = {Journal of Management Information Systems},
  volume = {26},
  number = {1},
  pages = {49--72},
  publisher = {Routledge},
  issn = {0742-1222},
  doi = {10.2753/MIS0742-1222260103},
  urldate = {2025-10-14}
}

@inproceedings{lockton2019NewMetaphorsWorkshop,
  title = {New {{Metaphors}}: {{A Workshop Method}} for {{Generating Ideas}} and {{Reframing Problems}} in {{Design}} and {{Beyond}}},
  shorttitle = {New {{Metaphors}}},
  booktitle = {Proceedings of the 2019 on {{Creativity}} and {{Cognition}}},
  author = {Lockton, Dan and Singh, Devika and Sabnis, Saloni and Chou, Michelle and Foley, Sarah and Pantoja, Alejandro},
  year = 2019,
  month = jun,
  pages = {319--332},
  publisher = {ACM},
  address = {San Diego CA USA},
  doi = {10.1145/3325480.3326570},
  urldate = {2025-11-17},
  isbn = {978-1-4503-5917-7},
  langid = {english}
}

@article{auger2013SpeculativeDesignCrafting,
  title = {Speculative Design: Crafting the Speculation},
  shorttitle = {Speculative Design},
  author = {Auger, James},
  year = 2013,
  month = mar,
  journal = {Digital Creativity},
  volume = {24},
  number = {1},
  pages = {11--35},
  issn = {1462-6268, 1744-3806},
  doi = {10.1080/14626268.2013.767276},
  urldate = {2026-01-23},
  langid = {english}
}

@article{burke1941four,
  title={Four master tropes},
  author={Burke, Kenneth},
  journal={The Kenyon Review},
  volume={3},
  number={4},
  pages={421--438},
  year={1941},
  publisher={JSTOR}
}

@book{agre1997computation,
  author = {Agre, Philip E.},
  title = {Computation and Human Experience},
  year = {1997},
  publisher = {Cambridge University Press},
  address = {Cambridge}
}

@incollection{schon1979generative,
  author = {Sch{\"o}n, Donald A.},
  title = {Generative Metaphor: A Perspective on Problem-Setting in Social Policy},
  booktitle = {Metaphor and Thought},
  editor = {Ortony, Andrew},
  year = {1979},
  publisher = {Cambridge University Press},
  address = {Cambridge},
}

@book{lakoff1980metaphors,
  author = {Lakoff, George and Johnson, Mark},
  title = {Metaphors We Live By},
  year = {1980},
  publisher = {University of Chicago Press},
  address = {Chicago}
}

@inproceedings{10.1145/3715336.3735714,
author = {Blythe, Mark and Lindley, Si\^{a}n and Murray-Rust, Dave},
title = {Artificial Intelligence and other Speculative Metaphors},
year = {2025},
isbn = {9798400714856},
publisher = {Association for Computing Machinery},
address = {New York, NY, USA},
url = {https://doi.org/10.1145/3715336.3735714},
doi = {10.1145/3715336.3735714},
booktitle = {Proceedings of the 2025 ACM Designing Interactive Systems Conference},
pages = {347–356},
numpages = {10},
location = {
},
series = {DIS '25}
}

@inproceedings{10.1145/3173574.3174194,
author = {Beck, Jordan and Ekbia, Hamid R.},
title = {The Theory-Practice Gap as Generative Metaphor},
year = {2018},
isbn = {9781450356206},
publisher = {Association for Computing Machinery},
address = {New York, NY, USA},
url = {https://doi.org/10.1145/3173574.3174194},
doi = {10.1145/3173574.3174194},
booktitle = {Proceedings of the 2018 CHI Conference on Human Factors in Computing Systems},
pages = {1–11},
numpages = {11},
location = {Montreal QC, Canada},
series = {CHI '18}
}

@misc{liang2025WidespreadAdoptionLarge,
  title = {The {{Widespread Adoption}} of {{Large Language Model-Assisted Writing Across Society}}},
  author = {Liang, Weixin and Zhang, Yaohui and Codreanu, Mihai and Wang, Jiayu and Cao, Hancheng and Zou, James},
  year = 2025,
  month = feb,
  number = {arXiv:2502.09747},
  eprint = {2502.09747},
  primaryclass = {cs},
  publisher = {arXiv},
  doi = {10.48550/arXiv.2502.09747},
  urldate = {2025-08-20},
  archiveprefix = {arXiv},
  langid = {english}
}

@article{ravselj2025HigherEducationStudents,
  title = {Higher Education Students' Perceptions of {{ChatGPT}}: {{A}} Global Study of Early Reactions},
  shorttitle = {Higher Education Students' Perceptions of {{ChatGPT}}},
  author = {Rav{\v s}elj, Dejan and Ker{\v z}i{\v c}, Damijana and Toma{\v z}evi{\v c}, Nina and Umek, Lan and Brezovar, Nejc and A. Iahad, Noorminshah and Abdulla, Ali Abdulla and Akopyan, Anait and Aldana Segura, Magdalena Waleska and AlHumaid, Jehan and Allam, Mohamed Farouk and All{\'o}, Maria and Andoh, Raphael Papa Kweku and Andronic, Octavian and Arthur, Yarhands Dissou and Ayd{\i}n, Fatih and Badran, Amira and {Balbont{\'i}n-Alvarado}, Roxana and Ben Saad, Helmi and Bencsik, Andrea and Benning, Isaac and Besimi, Adrian and Bezerra, Denilson Da Silva and Buizza, Chiara and Burro, Roberto and Bwalya, Anthony and Cachero, Cristina and {Castillo-Briceno}, Patricia and Castro, Harold and Chai, Ching Sing and Charalambous, Constadina and Chiu, Thomas K. F. and Clipa, Otilia and Colombari, Ruggero and Corral Escobedo, Luis Jos{\'e} H. and Costa, El{\'i}sio and Cre{\textcommabelow t}ulescu, Radu George and Crispino, Marta and Cucari, Nicola and Dalton, Fergus and Demir Kaya, Meva and {Dumi{\'c}-{\v C}ule}, Ivo and Dwidienawati, Diena and Ebardo, Ryan and Egbenya, Daniel Lawer and Faris, MoezAlIslam Ezzat and Fe{\v c}ko, Miroslav and Ferrinho, Paulo and Florea, Adrian and Fong, Chun Yuen and Francis, Zo{\"e} and Ghilardi, Alberto and {Gonz{\'a}lez-Fern{\'a}ndez}, Belinka and Hau, Daniela and Hossain, Md. Shamim and Hug, Theo and Inasius, Fany and Ismail, Maryam Jaffar and Jahi{\'c}, Hatid{\v z}a and Jessa, Morrison Omokiniovo and Kapanadze, Marika and Kar, Sujita Kumar and Kateeb, Elham Talib and Kaya, Feridun and Khadri, Hanaa Ouda and Kikuchi, Masao and Kobets, Vitaliy Mykolayovych and Kostova, Katerina Metodieva and Krasmane, Evita and Lau, Jesus and Law, Wai Him Crystal and Laz{\u a}r, Florin and {Lazovi{\'c}-Pita}, Lejla and Lee, Vivian Wing Yan and Li, Jingtai and {L{\'o}pez-Aguilar}, Diego Vinicio and Luca, Adrian and Luciano, Ruth Garcia and {Machin-Mastromatteo}, Juan D. and Madi, Marwa and Manguele, Alexandre Louren{\c c}o and Manrique, Rub{\'e}n Francisco and Mapulanga, Thumah and Marimon, Frederic and Marinova, Galia Ilieva and {Mas-Machuca}, Marta and {Mej{\'i}a-Rodr{\'i}guez}, Oliva and {Meletiou-Mavrotheris}, Maria and {M{\'e}ndez-Prado}, Silvia Mariela and {Meza-Cano}, Jos{\'e} Manuel and Mir{\c k}e, Evija and Mishra, Alpana and Mital, Ondrej and Mollica, Cristina and Morariu, Daniel Ionel and Mospan, Natalia and Mukuka, Angel and Navarro Jim{\'e}nez, Silvana Guadalupe and Nikaj, Irena and Nisheva, Maria Mihaylova and Nisiforou, Efi and Njiku, Joseph and Nomnian, Singhanat and {Nuredini-Mehmedi}, Lulzime and Nyamekye, Ernest and Obadi{\'c}, Alka and Okela, Abdelmohsen Hamed and {Olenik-Shemesh}, Dorit and Ostoj, Izabela and {Peralta-Rizzo}, Kevin Javier and Pe{\v s}tek, Almir and {Pilav-Veli{\'c}}, Amila and Pires, Dilma Rosanda Miranda and Rabin, Eyal and Raccanello, Daniela and Ramie, Agustine and Rashid, Md. Mamun Ur and Reuter, Robert A. P. and Reyes, Valentina and Rodrigues, Ana Sofia and Rodway, Paul and Ru{\v c}insk{\'a}, Silvia and Sadzaglishvili, Shorena and Salem, Ashraf Atta M. S. and Savi{\'c}, Gordana and Schepman, Astrid and Shahpo, Samia Mokhtar and Snouber, Abdelmajid and Soler, Emma and Sonyel, Bengi and Stefanova, Eliza and Stone, Anna and Strzelecki, Artur and Tanaka, Tetsuji and Tapia Cortes, Carolina and {Teira-Fachado}, Andrea and Tilga, Henri and Titko, Jelena and Tolmach, Maryna and Turmudi, Dedi and {Varela-Candamio}, Laura and Vekiri, Ioanna and Vicentini, Giada and Woyo, Erisher and Yorulmaz, {\"O}zlem and Yunus, Said A. S. and Zamfir, Ana-Maria and Zhou, Munyaradzi and Aristovnik, Aleksander},
  editor = {Wang, Chengliang},
  year = 2025,
  month = feb,
  journal = {PLOS ONE},
  volume = {20},
  number = {2},
  pages = {e0315011},
  publisher = {Public Library of Science (PLoS)},
  issn = {1932-6203},
  doi = {10.1371/journal.pone.0315011},
  urldate = {2025-08-05},
  copyright = {http://creativecommons.org/licenses/by/4.0/},
  langid = {english}
}

@article{asmelash2023thesebooksused,
  author  = "Asmelash, Leah",
  title   = "These books are being used to train AI. No one told the authors ",
  journal = "CNN",
  year = "2023",
  month = "Oct",
  url     = {https://www.cnn.com/2023/10/08/style/ai-books3-authors-nora-roberts-cec/index.html}
}

@misc{authorsguildsurvey,
  title = {New Authors Guild AI Survey Reveals That Authors Overwhelmingly Want Consent and Compensation for Use of Their Works},
  howpublished = {\url{https://authorsguild.org/news/ag-ai-survey-reveals-authors-overwhelmingly-want-consent-and-compensation-for-use-of-their-works/}},
  note = {Accessed: 2024-09-11}
}

@online{copyrightalliance2026,
  author       = {Madigan, Kevin},
  title        = {AI Copyright Lawsuit Developments in 2025: A Year in Review},
  organization = {Copyright Alliance},
  year         = {2026},
  month        = {January 8},
  url          = {https://copyrightalliance.org/ai-copyright-lawsuit-developments-2025/},
  note         = {Accessed 4 Feb.\ 2026}
}

@online{authorsguild2025anthropic,
  author       = {{The Authors Guild}},
  title        = {Bartz v. Anthropic Settlement: What Authors Need to Know},
  organization = {Authors Guild},
  year         = {2025},
  month        = {December 5},
  url          = {https://authorsguild.org/advocacy/artificial-intelligence/what-authors-need-to-know-about-the-anthropic-settlement/},
  note         = {Accessed 4 Feb.\ 2026}
}

@misc{kadrey_meta_order2025,
  title        = {Order in \textit{Kadrey v.\ Meta Platforms, Inc.}},
  howpublished = {U.S.\ District Court Order},
  author       = {{United States District Court}},
  year         = {2025},
  note         = {No.\ 3:23-cv-03417-VC, N.D.\ Cal.},
  url          = {https://media.npr.org/assets/artslife/arts/2025/order1.pdf}
}

@misc{bartz_anthropic_order2025,
  title        = {Order in \textit{Bartz v.\ Anthropic, Inc.}},
  howpublished = {U.S.\ District Court Order},
  author       = {{United States District Court}},
  year         = {2025},
  note         = {No.\ C 24-05417 WHA (N.C. of Cal.), PDF document},
  url          = {https://copyrightalliance.org/wp-content/uploads/2025/06/Bartz-v.-Anthropic-Order.pdf}
}

@inproceedings{longpre2024position,
  title={Position: Data Authenticity, Consent, \& Provenance for AI are all broken: what will it take to fix them?},
  author={Longpre, Shayne and Mahari, Robert and Obeng-Marnu, Naana and Brannon, William and South, Tobin and Gero, Katy Ilonka and Pentland, Alex and Kabbara, Jad},
  booktitle={Forty-first International Conference on Machine Learning},
  year={2024}
}

@online{bota_moisin_3cs2017,
  author       = {Monica Bo\c{t}a-Moisin},
  title        = {The 3Cs – Consent, Credit, Compensation},
  organization = {Cultural Intellectual Property Rights Initiative®},
  year         = {2017},
  url          = {https://www.culturalintellectualproperty.com/the-3cs},
  note         = {Accessed 4 Feb.\ 2026; framework for fair and equitable engagement with Indigenous Peoples, ethnic groups, and Local Communities}
}

@inproceedings{kyi2025GovernanceGenerativeAI,
  title = {Governance of {{Generative AI}} in {{Creative Work}}: {{Consent}}, {{Credit}}, {{Compensation}}, and {{Beyond}}},
  shorttitle = {Governance of {{Generative AI}} in {{Creative Work}}},
  booktitle = {Proceedings of the 2025 {{CHI Conference}} on {{Human Factors}} in {{Computing Systems}}},
  author = {Kyi, Lin and Mahuli, Amruta and Silberman, M. Six and Binns, Reuben and Zhao, Jun and Biega, Asia J.},
  year = 2025,
  month = apr,
  series = {{{CHI}} '25},
  pages = {1--16},
  publisher = {Association for Computing Machinery},
  address = {New York, NY, USA},
  doi = {10.1145/3706598.3713799},
  urldate = {2025-07-22},
  isbn = {979-8-4007-1394-1}
}

@misc{sharma2025PRAC3PrivacyReputation,
  title = {{{PRAC3}} ({{Privacy}}, {{Reputation}}, {{Accountability}}, {{Consent}}, {{Credit}}, {{Compensation}}): {{Long Tailed Risks}} of {{Voice Actors}} in {{AI Data-Economy}}},
  shorttitle = {{{PRAC3}} ({{Privacy}}, {{Reputation}}, {{Accountability}}, {{Consent}}, {{Credit}}, {{Compensation}})},
  author = {Sharma, Tanusree and Zhou, Yihao and Berisha, Visar},
  year = 2025,
  month = jul,
  number = {arXiv:2507.16247},
  eprint = {2507.16247},
  primaryclass = {cs},
  publisher = {arXiv},
  doi = {10.48550/arXiv.2507.16247},
  urldate = {2026-02-03},
  archiveprefix = {arXiv}
}

@article{braunstein_warren_2021,
  author       = {Braunstein, Laura and Warren, Michelle R.},
  title        = {Zombies in the Library Stacks},
  journal      = {Dartmouth Library Staff Publications},
  volume       = {41},
  year         = {2021},
  url          = {https://digitalcommons.dartmouth.edu/dlstaffpubs/41},
  note         = {Accessed 4 Feb.\ 2026}
}

@incollection{braun2012ThematicAnalysis,
  title = {Thematic Analysis.},
  booktitle = {{{APA}} Handbook of Research Methods in Psychology, {{Vol}} 2: {{Research}} Designs: {{Quantitative}}, Qualitative, Neuropsychological, and Biological.},
  author = {Braun, Virginia and Clarke, Victoria},
  editor = {Cooper, Harris and Camic, Paul M. and Long, Debra L. and Panter, A. T. and Rindskopf, David and Sher, Kenneth J.},
  year = 2012,
  pages = {57--71},
  publisher = {American Psychological Association},
  address = {Washington},
  doi = {10.1037/13620-004},
  urldate = {2025-09-03},
  isbn = {978-1-4338-1005-3},
  langid = {english}
}

@book{beaufort2007college,
  author    = {Beaufort, Anne},
  title     = {College Writing and Beyond: A New Framework for University Writing Instruction},
  publisher = {Utah State University Press},
  year      = {2007},
  address   = {Logan, UT},
  isbn      = {9780874216592},
  note      = {Imprint of University Press of Colorado},
  url       = {https://upcolorado.com/utah-state-university-press/college-writing-and-beyond}
}

@book{swales1990genre,
  author    = {Swales, John M.},
  title     = {Genre Analysis: English in Academic and Research Settings},
  publisher = {Cambridge University Press},
  year      = {1990},
  address   = {Cambridge, UK},
  isbn      = {9780521339778}
}

@article{couldry2019DataColonialismRethinking,
  title = {Data {{Colonialism}}: {{Rethinking Big Data}}'s {{Relation}} to the {{Contemporary Subject}}},
  shorttitle = {Data {{Colonialism}}},
  author = {Couldry, Nick and Mejias, Ulises A.},
  year = 2019,
  month = may,
  journal = {Television \& New Media},
  volume = {20},
  number = {4},
  pages = {336--349},
  issn = {1527-4764, 1552-8316},
  doi = {10.1177/1527476418796632},
  urldate = {2025-02-02},
  langid = {english}
}

@article{nplusone2025large,
  author       = {{The Editors}},
  title        = {Large Language Muddle},
  journal      = {n+1},
  year         = {2025},
  volume       = {Issue 51},
  number       = {Force Majeure},
  url          = {https://www.nplusonemag.com/issue-51/the-intellectual-situation/large-language-muddle/},
  note         = {Published Fall 2025},
}

@misc{ding2024MyVoiceYour,
  title = {My {{Voice}}, {{Your Voice}}, {{Our Voice}}: {{Attitudes Towards Collective Governance}} of a {{Choral AI Dataset}}},
  shorttitle = {My {{Voice}}, {{Your Voice}}, {{Our Voice}}},
  author = {Ding, Jennifer and J{\"a}ger, Eva and Ivanova, Victoria and Bunz, Mercedes},
  year = 2024,
  month = dec,
  number = {arXiv:2412.01433},
  eprint = {2412.01433},
  primaryclass = {cs},
  publisher = {arXiv},
  doi = {10.48550/arXiv.2412.01433},
  urldate = {2025-02-10},
  archiveprefix = {arXiv},
  langid = {english}
}

@inproceedings{jernite2022DataGovernanceAge,
  title = {Data {{Governance}} in the {{Age}} of {{Large-Scale Data-Driven Language Technology}}},
  booktitle = {2022 {{ACM Conference}} on {{Fairness}}, {{Accountability}}, and {{Transparency}}},
  author = {Jernite, Yacine and Nguyen, Huu and Biderman, Stella and Rogers, Anna and Masoud, Maraim and Danchev, Valentin and Tan, Samson and Luccioni, Alexandra Sasha and Subramani, Nishant and Johnson, Isaac and Dupont, Gerard and Dodge, Jesse and Lo, Kyle and Talat, Zeerak and Radev, Dragomir and Gokaslan, Aaron and Nikpoor, Somaieh and Henderson, Peter and Bommasani, Rishi and Mitchell, Margaret},
  year = 2022,
  month = jun,
  pages = {2206--2222},
  publisher = {ACM},
  address = {Seoul Republic of Korea},
  doi = {10.1145/3531146.3534637},
  urldate = {2025-02-02},
  isbn = {978-1-4503-9352-2},
  langid = {english}
}

@article{mio1997metaphor,
  title={Metaphor and politics},
  author={Mio, Jeffery Scott},
  journal={Metaphor and symbol},
  volume={12},
  number={2},
  pages={113--133},
  year={1997},
  publisher={Taylor \& Francis}
}

@online{latimer2026usecases,
  author       = {{FutureSum AI, Inc.}},
  title        = {Use Cases — Latimer.ai},
  year         = {2026},
  url          = {https://www.latimer.ai/#use-cases},
  note         = {Accessed: 2026-02-06},
  organization = {Latimer.ai},
}

@inproceedings{suresh2024ParticipationAgeFoundation,
  title = {Participation in the Age of Foundation Models},
  booktitle = {The 2024 {{ACM Conference}} on {{Fairness}}, {{Accountability}}, and {{Transparency}}},
  author = {Suresh, Harini and Tseng, Emily and Young, Meg and Gray, Mary and Pierson, Emma and Levy, Karen},
  year = 2024,
  month = jun,
  pages = {1609--1621},
  publisher = {ACM},
  address = {Rio de Janeiro Brazil},
  doi = {10.1145/3630106.3658992},
  urldate = {2025-02-02},
  isbn = {979-8-4007-0450-5},
  langid = {english}
}

@online{sloan2016writing,
  author       = {Sloan, Robin},
  title        = {Writing with the machine},
  year         = {2016},
  month        = {May},
  url          = {https://www.robinsloan.com/notes/writing-with-the-machine/},
  note         = {Accessed: 2026-02-06},
}

@misc{olmo2025olmo3,
      title={Olmo 3}, 
      author={Team Olmo and : and Allyson Ettinger and Amanda Bertsch and Bailey Kuehl and David Graham and David Heineman and Dirk Groeneveld and Faeze Brahman and Finbarr Timbers and Hamish Ivison and Jacob Morrison and Jake Poznanski and Kyle Lo and Luca Soldaini and Matt Jordan and Mayee Chen and Michael Noukhovitch and Nathan Lambert and Pete Walsh and Pradeep Dasigi and Robert Berry and Saumya Malik and Saurabh Shah and Scott Geng and Shane Arora and Shashank Gupta and Taira Anderson and Teng Xiao and Tyler Murray and Tyler Romero and Victoria Graf and Akari Asai and Akshita Bhagia and Alexander Wettig and Alisa Liu and Aman Rangapur and Chloe Anastasiades and Costa Huang and Dustin Schwenk and Harsh Trivedi and Ian Magnusson and Jaron Lochner and Jiacheng Liu and Lester James V. Miranda and Maarten Sap and Malia Morgan and Michael Schmitz and Michal Guerquin and Michael Wilson and Regan Huff and Ronan Le Bras and Rui Xin and Rulin Shao and Sam Skjonsberg and Shannon Zejiang Shen and Shuyue Stella Li and Tucker Wilde and Valentina Pyatkin and Will Merrill and Yapei Chang and Yuling Gu and Zhiyuan Zeng and Ashish Sabharwal and Luke Zettlemoyer and Pang Wei Koh and Ali Farhadi and Noah A. Smith and Hannaneh Hajishirzi},
      year={2025},
      eprint={2512.13961},
      archivePrefix={arXiv},
      primaryClass={cs.CL},
      url={https://arxiv.org/abs/2512.13961}, 
}

@article{lee2007BoundaryNegotiatingArtifacts,
  title = {Boundary {{Negotiating Artifacts}}: {{Unbinding}} the {{Routine}} of {{Boundary Objects}} and {{Embracing Chaos}} in {{Collaborative Work}}},
  shorttitle = {Boundary {{Negotiating Artifacts}}},
  author = {Lee, Charlotte P.},
  year = 2007,
  month = jun,
  journal = {Computer Supported Cooperative Work (CSCW)},
  volume = {16},
  number = {3},
  pages = {307--339},
  issn = {0925-9724, 1573-7551},
  doi = {10.1007/s10606-007-9044-5},
  urldate = {2026-02-06},
  langid = {english}
}

@article{szep2026FinetuningLargeLanguage,
  title = {Fine-Tuning {{Large Language Models}} with {{Limited Data}}: {{A Survey}} and {{Practical Guide}}},
  shorttitle = {Fine-Tuning {{Large Language Models}} with {{Limited Data}}},
  author = {Szep, Marton and Rueckert, Daniel and {Von Eisenhart-Rothe}, R{\"u}diger and Hinterwimmer, Florian},
  year = 2026,
  month = apr,
  journal = {Transactions of the Association for Computational Linguistics},
  volume = {14},
  pages = {341--377},
  issn = {2307-387X},
  doi = {10.1162/TACL.a.627},
  urldate = {2026-04-20},
  langid = {english}
}


\appendix
\clearpage

\section{Methodology}
\label{app:methodology}
\begin{table*}[!t]
\centering
\small
\renewcommand{\arraystretch}{1.25} 
\begin{tabular}{p{2cm} p{2cm} p{4.5cm} p{6cm}}
\toprule
\textbf{Starts at…} & \textbf{Duration} & \textbf{What happens} & \textbf{Who does what} \\
\midrule
0:00 & 5 min & Waiting for folks to arrive & Low key chatting with folks as they come in \\
0:05 & 10 min & Ground rules, tone setting & Leaders talk through slides; facilitators intro themselves \\
0:15 & 3 min & Participant introductions in the chat & Leaders go through slides \\
0:18 & 2 min & Agenda, questions before starting & Leaders go through slides \\
0:20 & 5 min & Warm-up question in chat & Leaders go through slides \\
0:25 & 10 min & Brainstorm model metaphors in Miro &
Facilitators and participants help with clustering; leaders low key summarize as it’s happening; facilitators start and encourage the clustering \\
0:35 & 20 min & People self-select into breakout rooms; we might help even out rooms by encouraging moving around &
Leaders create and manage breakout rooms; facilitators manage breakout rooms (at least one per room) \\
0:55 & 5 min & Break! &  \\
1:00 & 20 min & Reporting back & Leaders ask for group reporters \\
1:20 & 10 min & Closing & Leaders go through slides \\
\bottomrule
\end{tabular}
\caption{Workshop run of show}
\label{tab:runofshow}
\end{table*}

\subsection{Workshop Structure}
Each workshop lasted 90 minutes and had 1-2 leaders (authors) and 3-5 facilitators (authors and previous attendees). Previous attendees were invited to become facilitators for future workshops. 
\autoref{tab:runofshow} describes the “run of show” for each workshop with timing and what the leaders and facilitators were responsible for in each portion. 

\paragraph{Introduction and tone setting.} Workshops began with leaders introducing the goals of the workshop: how to structure a community language model. We went over what is a language model, what are metaphors for community models, and why we want to think about metaphors. We then went through an informed consent procedure where we described what was recorded and what was not, and had each facilitator introduce themselves. Finally, we had all participants optionally introduce themselves in the chat by sharing their name, where they were joining from, and a favorite technology. We then went over the agenda for the rest of the workshop.

\paragraph{Warm-up activity.} We asked participants to think of one to three words that described “How you would want an interaction with a language model to make you feel?” We had participants share these in the chat “waterfall” style: everyone typed in their response but didn't hit send until the leader said so. This allows the chat to “overflow” with everyone’s responses at once, and helped participants share their authentic thoughts without feeling influenced or overwhelmed by others. The leader briefly discussed what emerged from this activity.

\paragraph{Brainstorming the metaphors.} At this point we shared a link to a Miro board---a digital whiteboard that could be accessed without an account. Participants were asked to place sticky-notes on the whiteboard for each metaphor for community-governed models they could think of. We seeded this activity with the metaphors “food co-op”, “science experiment”, “library”, “seed bank”, and “community garden”. With the entire group, the leaders would discuss different metaphors that were being shared, sometimes asking the contributor to explain what they meant or add details if the text was ambiguous. Participants could add as many metaphors as they liked.

\paragraph{Clustering the metaphors.} As metaphor generation slowed, the facilitators started to cluster the metaphors by dragging similar metaphors close to each other. After participants finished adding metaphors, they joined in on the clustering. No specific criteria were given for how to cluster, rather this was meant as a quick way to surface potential groups of interest. \autoref{fig:miroboardclusters} shows some example clusters from the workshops.

\paragraph{Break-out rooms for metaphor analysis.} Depending on how many clusters emerged, breakout rooms were created for each cluster. (We aimed to have enough breakout rooms such that each room had a minimum of 3 participants. Some clusters were merged if the cluster-to-participant ratio was too high.) Participants then self-selected into the breakout room for the cluster they were most interested in, and a facilitator joined each room. Each room was asked to analyze one or two metaphors from their cluster in detail. We provided a worksheet in the Miro board that had spaces for participants to answer questions about a specific metaphor in detail. We show an example worksheet, filled in by participants, in \autoref{fig:exampleworksheet}. Each group had at least one facilitator present. The facilitator would help participants understand the goals of the activity (to map the metaphor to how governance of an LLM might work) as well as the intention behind the questions in the workshop. Facilitators also encouraged all members of the group to speak or contribute if they liked, helped participants stay on track, and would take notes if need be on the Miro board.

\paragraph{Reporting back.} After a brief break, each group had a few minutes to report back about what they discussed in their group. 

\paragraph{Closing.} The leaders would sum up interesting points that had emerged, then talk about the next workshops: they encouraged people to invite others, or attend as a facilitator, and to join a group Discord to stay up to date with what happens next in the project. In the final workshop, we hosted a group discussion about next steps for the project and how people would like to be involved.

\section{Metaphor dimension: Structure}
\label{app:structure}

In addition to domain (Section \ref{sectionfour}), we analyzed metaphors along a structural dimension that captures the form participants used to conceptualize language models. \autoref{tab:structure_metaphors} summarizes the five structures we identified—objects, places, groups, processes, and systems—with definitions and example metaphors from the workshops.

\begin{table*}[!t]
\centering
\renewcommand{\arraystretch}{1.25}
\begin{tabular}{p{1.5cm}p{6cm}p{7cm}}
\toprule
\textbf{Structure} & \textbf{Definition} & \textbf{Example metaphors from workshop} \\
\midrule
Object & A material thing to be seen and touched & easel, bbq, loom, aquarium, little free public library, compost pile \\
\midrule
Place & A space available or designated for a purpose & coffee shop, public library, farmer’s market, bike kitchen, village square, community garden, monastery \\
\midrule
Group & A collection of people or things classed together & labor union, hiking club, buy nothing group, seed swap, meerkat colony, beehive, priesthood, group text thread \\
\midrule
Process & A series of actions or steps to achieve an end & barn raising, music performance, dice game, potluck \\
\midrule
System & A set of things or procedures working together as part of an interconnected network & wikipedia, NYC mesh, solar microgrids, ant colony, fungal network \\
\bottomrule
\end{tabular}
\caption{Metaphors from all four workshops were also analyzed along a structure dimension, capturing the forms participants used to conceptualize language models such as objects, places, groups, processes, and systems.}
\label{tab:structure_metaphors}
\end{table*}


\end{document}